\documentclass{aa}

\usepackage{graphicx}
\usepackage[T1]{fontenc}
\usepackage{ae,aecompl}
\usepackage{amssymb}
\usepackage{amsmath}
\usepackage{rotating, graphicx}
\usepackage{subfig}
\usepackage{pdflscape}
\usepackage{natbib}
\usepackage{amsmath}
\usepackage{courier}
\usepackage{color}
\usepackage{float}
\restylefloat{table}
\usepackage{footnote}
\usepackage[T1]{fontenc}
\usepackage{aecompl}
\usepackage{amssymb}
\usepackage{rotating}
\usepackage{mathtools}
\usepackage{enumitem,booktabs,cfr-lm}
\usepackage[referable]{threeparttablex}
\renewlist{tablenotes}{enumerate}{1}
\usepackage{multirow}

\makeatletter
\setlist[tablenotes]{label=\tnote{\alph*},ref=\alph*,itemsep=\z@,topsep=\z@skip,partopsep=\z@skip,parsep=\z@,itemindent=\z@,labelindent=\tabcolsep,labelsep=.2em,leftmargin=*,align=left,before={\footnotesize}}
\makeatother

\def\aap{Astronomy \& Astrophysics}

\def\apjs{The Astrophysical Journal Supplement}
\def\apjl{ApJL}

\usepackage{txfonts}
\begin{document}

   \title{Wide field-of-view study of the Eagle Nebula with the Fourier transform imaging spectrograph SITELLE at CFHT}

   \subtitle{}
   \titlerunning{Spectro-imaging study of the Eagle Nebula with SITELLE}
    
   \author{N. Flagey\thanks{Based on observations obtained at the Canada-France-Hawaii Telescope (CFHT) which is operated from the summit of Maunakea by the National Research Council (NRC) of Canada, the Institut National des Sciences de l'Univers (INSU) of the Centre National de la Recherche Scientifique (CNRS) of France, and the University of Hawaii. The observations at the Canada-France-Hawaii Telescope were performed with care and respect from the summit of Maunakea which is a significant cultural and historic site.}
          \inst{1}
          \and 
          A. F. McLeod,\inst{2,3}\\
          L. Aguilar\inst{1,4,5}
          \and
          S. Prunet\inst{1}
          }

   \institute{Canada-France-Hawaii Telescope Corporation, 65-1238 Mamalahoa Hwy, Kamuela, HI 96743, USA\\
   \email{flagey@cfht.hawaii.edu}
   \and
   Department of Astronomy, University of California Berkeley, Berkeley, CA 94720, USA\\
   \email{anna.mcleod@berkeley.edu}
   \and
   Department of Physics \& Astronomy, Texas Tech University, PO Box 41051, Lubbock, TX 79409, USA
   \and
   Universit\'e Pierre et Marie Curie, 4 Place Jussieu, 75005 Paris, France
   \and
   University College London, Department of Earth Sciences, Gower St, Bloomsbury, London WC1E 6BT, UK
   }

   \date{Received September 15, 1996; accepted March 16, 1997}

 
  \abstract
   {We present the very first wide-field, 11\arcmin\ by 11\arcmin, optical spectral mapping of M~16, one of the most famous star-forming regions in the Galaxy. The data were acquired with the new imaging Fourier transform spectrograph SITELLE mounted on the Canada-France-Hawaii Telescope (CFHT). We obtained three spectral cubes with a resolving power of 10'000 (SN1 filter), 1500 (SN2 filter) and 600 (SN3 filter), centered on the iconic Pillars of Creation and the HH~216 flow, covering the main optical nebular emission lines, namely [\ion{O}{ii}]$\lambda$3726,29 (SN1), H$\beta$, [\ion{O}{iii}]$\lambda$4959,5007 (SN2), [\ion{N}{ii}]$\lambda$6548,84, H$\alpha$, and [\ion{S}{ii}]$\lambda$6717,31 (SN3). }
   {We validate the performance, calibration, and data reduction of SITELLE, and analyze the structures in the large field-of-view in terms of their kinematics and nebular emission.}
   {We compared the SITELLE data to MUSE integral field observations and other spectroscopic and narrow-band imaging data to validate the performance of SITELLE. We computed gas-phase metallicities via the strong-line method, performed a pixel-by-pixel fit to the main emission lines to derive kinematics of the ionized gas, computed the mass-loss rate of the Eastern pillar (also known as the Spire), and combined the SITELLE data with near-infrared narrow-band imaging to characterize the HH~216 flow.}
   {The comparison with previously published fluxes demonstrates very good agreement. We disentangle the dependence of the gas-phase metallicities (derived via abundance-tracing line ratios) on the degree of ionization and obtain metallicities that are in excellent agreement with the literature. We confirm the bipolar structure of HH~216, find evidence for episodic accretion from the source of the flow, and identify its likely driving source. We compute the mass-loss rate $\dot{M}$ of the Spire pillar on the East side of the H{\sc ii} region and find excellent agreement with the correlation between the mass-loss rate and the ionizing photon flux from the nearby cluster NGC 6611.}
   {}
   \keywords{H{\sc ii} regions -- ISM: jets and outflows -- ISM: individual objects (M 16), individual objects (HH 216)
               }

   \maketitle
%

\section{Introduction}
With the ever increasing need of simultaneous large spatial and spectral coverage of astrophysical targets, integral field spectroscopy (IFS) has seen a rapid increase in development over the last decade. Among the advantages over conventional spectroscopy and (narrow- and broadband) imaging is the possibility of deriving physical properties and parameters, as well as kinematics, for the entire field-of-view (FOV) covered by the instrument. IFS, however, typically comes with the compromise of choosing between a large FOV and medium to low spectral resolution or a small FOV with high resolving power. The IFS capability of the optical Fourier-transform spectrograph SITELLE \citep{Grandmont12} offers the possibility of combining a large FOV with a range of resolving powers in the visible wavelength regime. This makes SITELLE an ideal instrument to analyze entire H{\sc ii} regions, where the stellar content drives the kinematics and properties of the surrounding gas. 

The Eagle Nebula (M 16) is one of the most famous and best-studied star-forming regions in the sky. The massive stars of the central cluster NGC 6611 are driving the expansion of the H{\sc ii} region and are responsible for shaping the surrounding cloud into iconic, ionized pillar-like structures protruding into the H{\sc ii} region. Because of this, M 16 is an ideal target when it comes to classical H{\sc ii} regions and feedback from massive stars. Furthermore, because this region has been studied in many wavelength ranges over the past decades (\citealt{hester96}, \citealt{white99}, \citealt{mccaugh02}, \citealt{andersen04}, \citealt{healy04}, \citealt{indeb07}, \citealt{flagey11}, \citealt{m16}, just to name a few), it is also a perfect testing ground for the performance and calibration of new instruments.

The stellar population of the central cluster, NGC 6611, is well studied and contains at least 13 O-type stars \citep{evans05}, of which the most massive one is HD 168076, an O4 III star \citep{sota11} of about 75-80 M$_{\odot}$ \citep{hillenbrand93}. It also contains at least 30 B-type stars and a significant population of intermediate- and low-mass stars \citep{belikov00}. The age of the cluster is of the order of 2-3 Myr and here we adopt a heliocentric distance of 2 kpc \citep{hillenbrand93}. 

The pillars and the surrounding H{\sc ii} region have been studied across almost the entire spectrum. \citet{linsky07} presented the Chandra X-ray data of the region and found that among the identified X-ray sources (predominantly consisting of low-mass pre-main-sequence and high-mass main sequence stars) some correspond to young stellar objects (YSOs) which are deeply embedded in the pillars, therefore presenting evidence for ongoing star formation within the pillar material. \cite{hester96} presented optical Hubble Space Telescope (HST) images of the central pillars, revealing a stratified ionization structure of the interface between pillar material and H{\sc ii} region, a photo-evaporative flow due to the intense ionizing radiation from NGC 6611, and a population of star-forming, cometary-shaped globules emerging from the pillars as the latter are being photo-evaporated. \cite{andersen04} performed a comprehensive study of the Herbig-Haro object HH 216 \citep[discovered and named M16 HH~1 by][]{Meaburn1982} at the base of the central pillars with data ranging from the millimeter to the optical, finding that HH 216 is the terminal bow shock of an HH flow with a jet and discussing a possible source for the latter. With molecular line data, \cite{white99} found that the pillar tips are dense cores which contain most of the pillars' mass, and also correspond to cool ($\sim$ 20~K) continuum cores. \cite{white99} also discuss the formation scenario of the pillars, in which dense cores of gas shielded the material behind them from the irradiation of the nearby massive stars, causing the pillar-like shape to be eroded from the surrounding molecular clouds. \citet{flagey11} analyzed the mid-infrared structure of the entire region and used a dust model to characterize the dust grain size distributions in the H{\sc ii} region and the surrounding photo-dissociation region (PDR) and constrain the energetics from the star cluster.

More recently, \citet{m16} (henceforth referred to as MC15) revisited the central pillar-like structures of M 16 with a 3\arcmin$\times$3\arcmin\ mosaic obtained with the optical integral field spectrograph MUSE \citep{muse} mounted on the VLT. MC15 exploited the large wavelength coverage of MUSE ($\sim$4750 - 9350 \AA) to obtain integrated line maps of the main nebular emission lines and compute an extinction as well as electron density and temperature maps. The detailed analysis of the ionization structure of the PDR at the pillar tips was compared to that of radiation transfer post-processed simulated pillar-like structures \citep{ercolano12}, finding a very good agreement between simulated and observed ionized pillars. MC15 computed a velocity map of the entire 3\arcmin$\times$3\arcmin\ mosaic and, together with an estimated mass of 200 M$_{\odot}$ \citep{white99}, computed a mass-loss rate of the pillars due to photo-evaporation from the nearby OB stars of $\approx$ 70 M$_{\odot}$ Myr$^{-1}$, and hence a pillar lifetime of about 3 Myr. Furthermore, MC15 identified HH 1176, a Herbig-Haro flow from a young stellar object which formed at the tip of the middle pillar.

Here, we present data from the new Fourier-transform spectrograph SITELLE mounted at the Cassegrain focus of CFHT on Maunakea. The data, spanning a single 11\arcmin$\times$11\arcmin\ FOV, covers the central pillars as well as a large part of the H{\sc ii} region and the surrounding clouds, including the north-eastern pillar known as the Spire (see Figure~\ref{rgb}), making this the largest combination of imaging and spectroscopy of the region to date. The discussed data is presented in Section \ref{sec:obs}. With SITELLE, we cover the main optical nebular emission lines, which allow us to perform a systematic performance analysis of the instrument by comparing the data to previously published measurements including MUSE integral field data in Section \ref{sec:results}. We compute line ratio maps and derive gas-phase metallicities in Section \ref{sec:diag}. In combination with near-IR data, we analyze the counter flow of HH~216 and identify the likely driving source in Section \ref{sec:hh}. We compute the mass-loss rate of the Spire and discuss the result in the context of the relation between the mass loss-rate and the ionizing photon-flux from massive stars in Section \ref{sec:spire}. A summary of the findings is presented in Section \ref{sec:ccl}.


\section{Observations}
\label{sec:obs}

The data were obtained in queue mode on July 10, 2016 and July 1, 2017 at the Canada-France-Hawaii Telescope on Maunakea, Hawaii. The observations were carried out with the new optical, dual port, large field of view (11\arcmin\ by 11\arcmin), imaging Fourier transform spectrograph (FTS) SITELLE \citep{Grandmont12}. In this paper we also use CFHT/WIRCam observations of the Eagle Nebula, obtained in 2008 with the Br$\gamma$, H$_2$ v=1-0 S(1), and $K_s$ filters and reduced by TeraPix \citep[e.g.,][]{Marmo07}.

\begin{figure*}[!t]
\centering
\includegraphics[width=\linewidth]{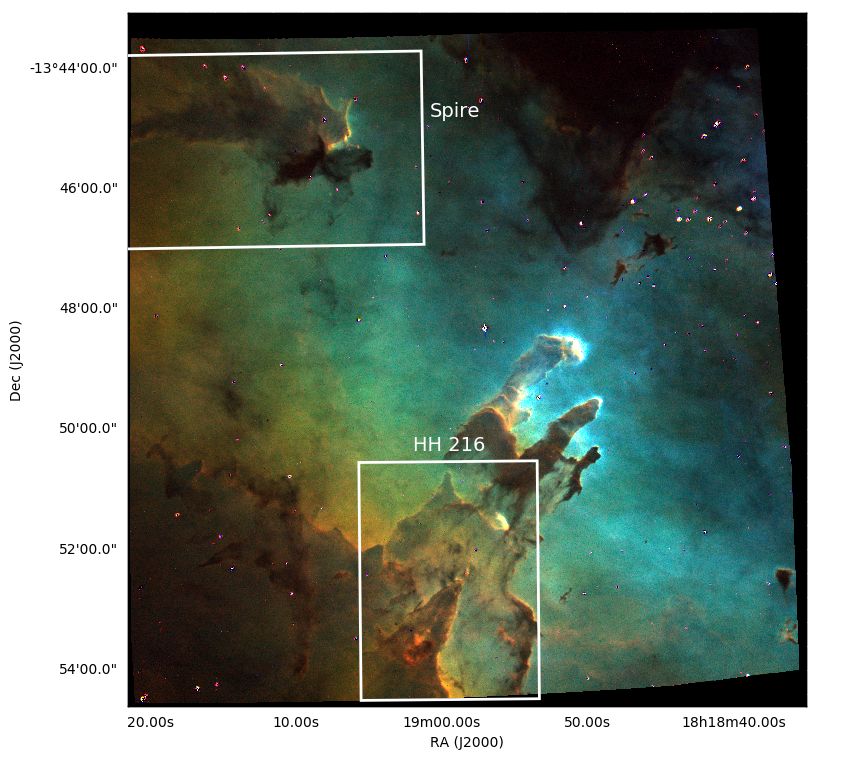}
\caption{Three-color continuum-subtracted and reddening-corrected composite (red = [\ion{S}{ii}]$\lambda$6717, green = H$\alpha$, blue = [\ion{O}{iii}]$\lambda$5007). The white rectangles mark the regions used to crop the large-scale field, centering on relevant structures, i.e. the pillar-like structure known as the Spire and the bright Herbig-Haro object known as HH~216 just south of the iconic Pillars of creation. In this projection, North is up and East is left.}
\label{rgb}
\end{figure*}

The data obtained with SITELLE at the telescope are interferogram cubes, where each frame corresponds to a given optical path difference (OPD) between the two mirrors in the interferometer. Multiple OPDs are scanned, with the mobile mirror moved by a fixed amount for each new frame. At each step, a new 2D interferogram is obtained on each of the two cameras. The interferogram cubes have been reduced by the SITELLE instrument team at CFHT and Universit\'e of Laval using ORBS\footnote{ \url{http://www.cfht.hawaii.edu/Instruments/Sitelle/SITELLE_orbs.php} }. The reduction process is a Fourier transform at its core, though several other critical operations are performed, including astrometric correction, camera alignment, phase correction, wavelength calibration, and flux calibration. The resulting spectral cubes have 2048 by 2064 spatial pixels and a number of spectral frames that depends on the resolution of the observations and the filter in use.

The observations we present in this paper use the SN1, SN2, and SN3 filters of SITELLE, which cover the spectral ranges from 365 to 385 nm, from 484 to 512 nm, and from 648 to 685 nm, respectively\footnote{The transmission curves of the interference filters are available at \url{https://www.cfht.hawaii.edu/Instruments/Sitelle/SITELLE_filters.php}}. Observations with SITELLE can use a spectral resolution that is adapted to the needs of the user by setting the maximum OPD. For M~16 we requested a resolving power of 10'000 with the SN1 filter, 600 with the SN2 filter, and 1500 with the SN3 filter. Our observations thus required 1714 interferogram steps with a 2.6 second integration time per step for the SN1 filter, 135 steps with a 15 second integration time per step with the SN2 filter, and 254 teps with a 3 second integration time per step with the SN3 filter. The image quality measured during the observations was $\sim$ 1.4\arcsec, 1.1\arcsec, and 1.1\arcsec\ respectively with the SN1, SN2, and SN3, respectively.

The processing of extended, bright emission line regions observed with SITELLE requires additional calibration data, in comparison with other astronomical targets. This is due to the absence of any detectable continuum in the M~16 spectra and, therefore, no possibility to estimate phase correction maps directly from the data. In the processing of FTS data, these phase correction maps are necessary to recover the exact location of the Zero Path Difference (ZPD) location, as a function of frequency, with respect to the optical path sampling of a data set, and to recover the pure cosine nature of the interferograms with respect to the frequency dependent ZPD.

\begin{figure*}[!t]
\centering
\includegraphics[width=0.33\linewidth]{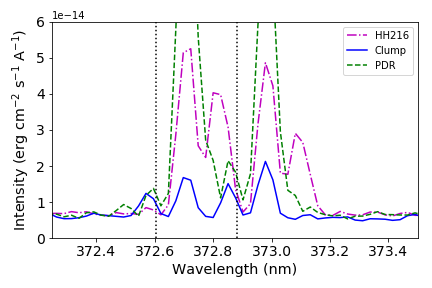}
\includegraphics[width=0.33\linewidth]{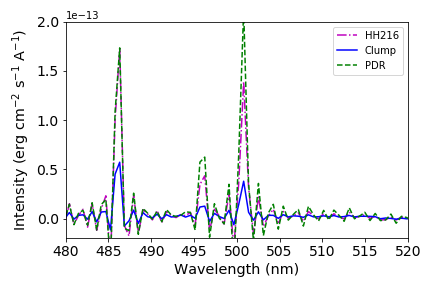}
\includegraphics[width=0.33\linewidth]{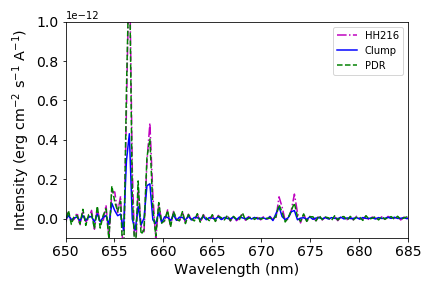}
\caption{Typical spectra observed in the SN1, SN2, and SN3 filters (top to bottom) of the SITELLE data. The whole spectral range is shown for the SN2 and SN3 filters while only a 1.25~nm range is shown for the SN1 filter to focus on the [\ion{O}{ii}] doublet and its multiple radial velocity components. For the SN1 filter, the reference wavelengths of the [\ion{O}{ii}] doublet are indicated by dotted lines. The spectra were obtained using apertures of 2.5\arcsec\ in diameter for SN1 and 13\arcsec\ in diameter for SN2 and SN3, at the locations of HH-216, a blueshifted clump (see section \ref{sec:vel}), and at the tip of Pillar 2.}
\label{fig:spectra}
\end{figure*}

To remedy this situation, additional calibration data are taken using continuum sources (either incandescence light bulbs or LEDs) to provide external phase correction maps. These additional calibrations were available for the SN2 and SN3 filters only. We therefore had to process the SN1 data in power spectrum: no phase correction was attempted, but the power from sine and cosine modes were added in quadrature. This results in a significant reduction in the spectral resolution of that cube (see section \ref{sec:sens}). However, the final resolution in SN1 still allows a good separation of the [\ion{O}{ii}] doublet (see Figure \ref{fig:spectra}).

Finally, because of the strength of the emission lines in M16, short exposure times are needed to avoid saturation in the CCDs. This in turn results in a very low signal-to-noise ratio (S/N) in the sky lines that are usually easy to measure, for instance in the SN3 band. Therefore these sky lines cannot be used for the wavelength calibration and one has to rely solely on external wavelength calibration maps obtained from dedicated green (HeNe) laser spectral cubes. The laser calibration is not as accurate and leaves a non negligible bias. Throughout this paper we will thus focus on relative radial velocities rather than absolute radial velocities.

Figure \ref{rgb} shows the 11\arcmin\ by 11\arcmin\ field of view covered by SITELLE with a commonly used color code: red is the [\ion{S}{ii}]$\lambda$6717 emission line, green is H$\alpha$, and blue is the [\ion{O}{iii}]$\lambda$5007 line. The iconic Pillars of Creation are at the center of the image, with the exciting star cluster NGC~6611 towards the top right corner. In the top left corner another pillar - the Spire - can be seen. The dark molecular cloud, parent of NGC~6611, is visible at the top right and bottom left of the image. Besides the stars (almost absent from the continuum-subtracted image in Figure \ref{rgb}), the brightest features in the image are the ionized tips of the Pillars and the Spire, the many PDRs, the ionized region within the Pillars, and a unique feature -- HH~216 -- located near the bottom center of the field.


\section{Emission line and radial velocity maps}
\label{sec:results}

Typical spectra observed with SITELLE towards different regions in M~16 (the PDR at the pillar tips, the bow shock of HH 216, and a blueshifted gas globule, see sections \ref{sec:vel} and \ref{sec:hh}) are shown in Figure~\ref{fig:spectra}. These reveal that the line spread function is close to a $sinc$ profile for the SN2 and SN3 filters. This is due to the Fourier transform of a truncated interferogram, which spans a finite range of optical path differences. An infinitely narrow emission line would lead to a $cosine$ function interferogram. Because the interferogram is sampled only between two maximum OPDs, it is a truncated $cosine$ interferogram, whose inverse Fourier transform leads to a $sinc$ emission line profile\footnote{For more details about FTS we invite the reader to consult the FTS primer written for SITELLE at \url{http://www.cfht.hawaii.edu/Instruments/Sitelle/FTS-primer-Jan16.pdf}}. It is thus possible to obtain spectra with negative values, on pixels where the continuum is weak, though this is purely due to the mathematical process required to convert the interferogram into a spectrum. The SN1 spectral cube that we analyze in this paper has gone through a different pipeline (see section \ref{sec:obs}), and the line profile are therefore that of a $\sqrt(sinc^2)$ instead. All the spectral cubes are in units of $F_\lambda$ as a function of wavenumber, since the OPD is in units of length. 

Several emission lines can be identified (see Table \ref{ref_wave}): the [\ion{O}{ii}] doublet at 3726/3729 {\AA} in the SN1 filter, the H$\beta$ line and the [\ion{O}{iii}] doublet at 4960/5007 {\AA} in the SN2 filter, and the H$\alpha$ line, the [\ion{N}{ii}] doublet at 6548/6584 {\AA}, and the [\ion{S}{ii}] doublet at 6717/6731 {\AA} in the SN3 filter. Other much fainter lines are also detected, like the He I line at 6678 {\AA}, but are not included in this paper. In addition, multiple radial velocity components of the [\ion{O}{ii}] lines are detected towards HH~216: one main component, one blueshifted, and one redshifted, with a separation of about 7~cm$^{-1}$ or 70~km/s.

\begin{table}[b]
  \centering
  \caption{Reference wavelengths (in {\AA}) used for the radial velocity maps. Source: the Atomic Line List v2.05b21 \citep{vanHoof2018}.}
  \begin{tabular}{l l r}
    \hline
    \multicolumn{2}{l}{Line}     & Wavelength (\AA)\\
    \hline
    \multirow{2}{*}{SN1} & [\ion{O}{II}]$\lambda$3726 & 3726.032 \\
    & [\ion{O}{II}]$\lambda$3729 & 3728.815 \\
    \hline
    \multirow{3}{*}{SN2} & H$\beta$ & 4861.325 \\
    & [\ion{O}{III}]$\lambda$4960   & 4958.911 \\
    & [\ion{O}{III}]$\lambda$5007   & 5006.843 \\
    \hline
	\multirow{5}{*}{SN3} & [\ion{N}{II}]$\lambda$6584  & 6548.04 \\
    & H$\alpha$                   & 6562.80 \\
    & [\ion{N}{II}]$\lambda$6548  & 6583.46 \\
    & [\ion{S}{II}]$\lambda$6717  & 6716.44 \\
    & [\ion{S}{II}]$\lambda$6731  & 6730.82 \\
    \hline
  \end{tabular}
  \label{ref_wave}
\end{table}

To extract lines fluxes and other relevant parameters from the spectral cubes, we fit each of the four million spectra (2048 by 2064 pixels) with a sum of $sinc$ profiles for the SN2 and SN3 spectral cubes and with a sum of $sinc^2$ profiles for the squared SN1 spectral cube, using the MPFIT package for IDL\footnote{ \url{https://www.physics.wisc.edu/~craigm/idl/fitting.html}}. A flat continuum is added to the sum of line profiles. For the SN3 and SN2 spectral cubes, we use one $sinc$ profile per emission line, since the spectral resolution does not allow us to disentangle multiple radial velocity components. For the SN1 spectral cube, the spectral resolution enables us to distinguish multiple radial velocity components for the [\ion{O}{ii}] doublet, and we thus define three $sinc^2$ profiles for each [\ion{O}{ii}] line: a main component, a blueshifted component, and a redshifted component. All the parameters used for the fit are given in Table \ref{tab:fit} in the Appendix. The initial values and the limits to the ranges that each parameter can explore are also provided. Since the fit to the spectral lines is performed on the spectral cubes, positions and widths are given in cm$^{-1}$.

\begin{figure*}[!t]
\centering
\subfloat[H$\alpha$]{\includegraphics[scale=0.38]{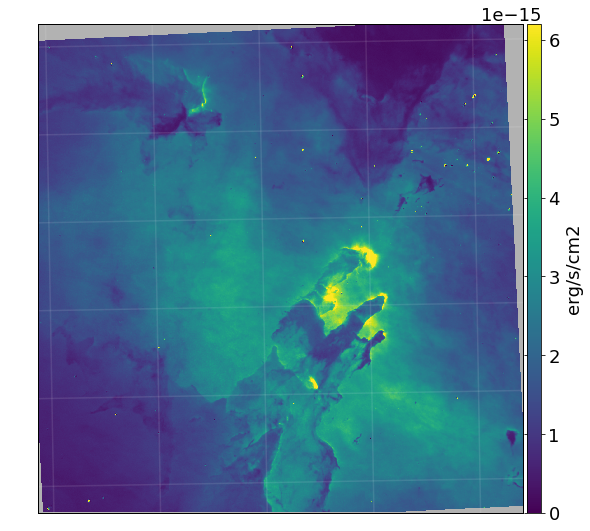}}
\hspace{0.05in}
\subfloat[ $\rm{[\ion{O}{iii}]\lambda 5007}$ ]{\includegraphics[scale=0.38]{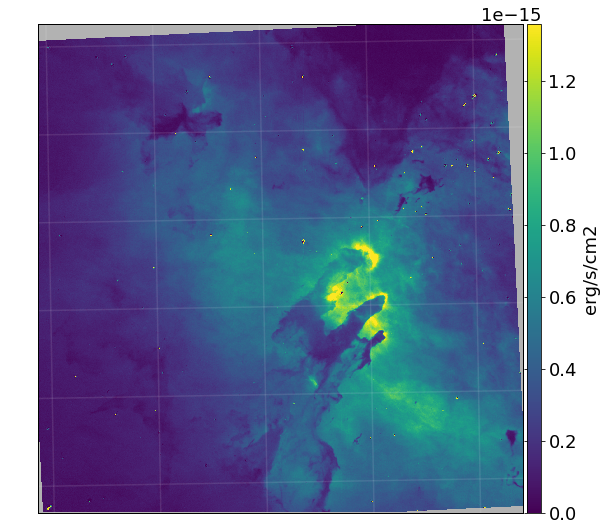}}\\
\subfloat[ $\rm{[\ion{S}{ii}]\lambda 6717}$ ]{\includegraphics[scale=0.38]{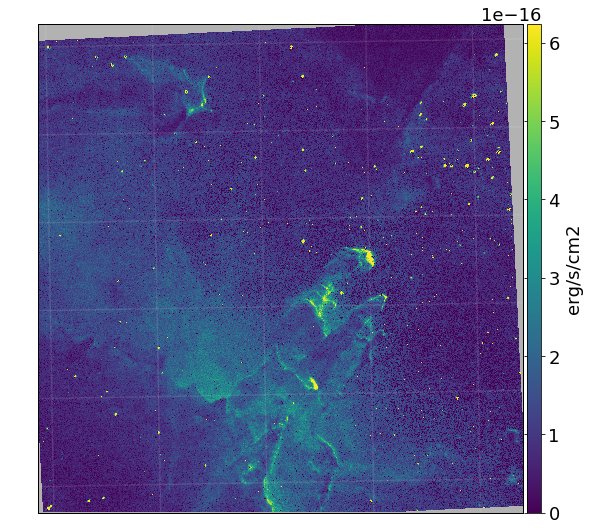}}
\hspace{0.05in}
\subfloat[ $\rm{[\ion{N}{ii}]\lambda 6584}$ ]{\includegraphics[scale=0.38]{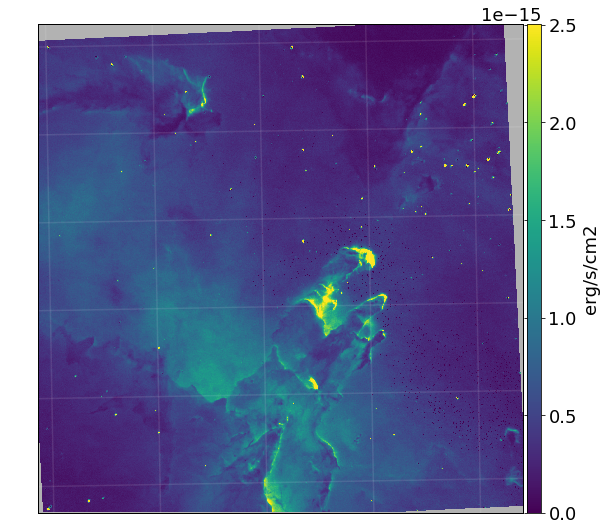}}
\caption{Emission line flux maps (in $\rm{erg/s/cm^2}$) for the H$\alpha$, the [\ion{O}{iii}]$\lambda 5007$, the [\ion{N}{ii}]$\lambda 6584$, and [\ion{S}{ii}]$\lambda 6717$ lines.}
\label{fig:fluxmaps}
\end{figure*}

The uncertainties in the spectra are inferred from the standard deviation of each spectrum, at the edges of the spectral coverage for each filter. This is because, as in any FTS and unlike in dispersive spectrographs, all the light passing through the filter contributes to the noise of every spectral element, which may be the dominant component for a source like M16. For the SN2 and SN3 filters, we chose wavelengths outside of the 645-690~nm and 480-515~nm range, respectively. This spectral range lies outside the filter coverage and is produced during the data reduction and only contains noise, which is independent of wavelength in a FTS. For the SN1 filter, we select the ranges from 372 to 372.5~nm and from 373.2 to 374~nm to estimate the standard deviation of each spectrum.

Thus, the results from the fit are maps for the centroids, amplitudes, and widths of each emission line, with their associated uncertainties, in addition to a continuum map for each filter. Flux maps are derived from the width and amplitude maps using Equation \ref{eq:sinc_area}:

\begin{equation}
  flux = \pi \times amplitude \times width
  \label{eq:sinc_area}
\end{equation}

which gives the integrated flux of a $sinc$ profile. Because the SN1 maps where obtained with a power spectrum method, we run the fit on the squared data cube, thus using $sinc^2$ functions, which have the same integral as a $sinc$ function and lead to the square of the flux maps. The centroid maps are converted into radial velocity maps using the reference wavelengths listed in Table \ref{ref_wave}.

Flux uncertainties maps are derived from the uncertainties maps on the amplitudes and widths of the emission lines, added in quadrature. The resulting uncertainties are typically lower than 5\% for the H$\alpha$ emission line, 5\% for the H$\beta$ line, 5-10\% for the [\ion{O}{iii}] lines, and 10-15\% for the [\ion{N}{ii}] lines. They are significantly larger for the weaker lines of [\ion{O}{ii}] and [\ion{S}{ii}].

\begin{table*}[t]
\centering
  \begin{tabular}{c c c c c c c c l}
    \hline
    \hline
    RA & Dec & Diam & [\ion{O}{ii}]$\lambda$3726 & [\ion{O}{ii}]$\lambda$3729 & H$\beta$ & [\ion{O}{iii}]$\lambda$5007 & H$\alpha$ & Ref\\
    \hline    
    18 15 57.9 & -13 54 13 & 3\arcmin11\arcsec & 0.497 & 0.683 & 1.55 & 1.81 & 8.51 & $^{a}$ \\
    18 18 48.2 & -13 52 57 & 3\arcmin11\arcsec & 0.499-0.659 & 0.589-0.723 & 1.4$\pm$0.3 & 1.61$\pm$0.05 & 8.2$\pm$0.2 & this paper \\
    \hline
    18 16 02.6 & -13 51 59 & 3\arcmin11\arcsec & 0.627 & 0.778 & 1.56 & 1.90 & 8.69 & $^{a}$ \\
    18 18 52.8 & -13 50 43 & 3\arcmin11\arcsec & 0.574-0.741 & 0.645-0.794 & 1.4$\pm$0.5 & 1.65$\pm$0.07 & 8.6$\pm$0.2 & this paper \\
    \hline
    18 16 02.6 & -13 51 59 & 1\arcmin35\arcsec & 0.201 & 0.254 & 0.405 & 0.444 & 2.44 & $^{a}$ \\
    18 18 52.8 & -13 50 43 & 1\arcmin35\arcsec & 0.148-0.182 & 0.166-0.195 & 0.3$\pm$0.1 & 0.32$\pm$0.01 & 1.87$\pm$0.05 & this paper \\
    18 18 52.8 & -13 50 43 & 1\arcmin35\arcsec & - & - & 0.285 & 0.346 & 1.89 & MC15 \\
    \hline
    18 16 14.9 & -13 50 38 & 1\arcmin35\arcsec & 0.173 & 0.229 & 0.449 & 0.402 & 2.56 & $^{a}$ \\
    18 19 05.1 & -13 49 21 & 3\arcmin11\arcsec & 0.146-0.178 & 0.168-0.196 & 0.36$\pm$0.08 & 0.32$\pm$0.01 & 2.17$\pm$0.05 & this paper \\
    \hline
  \end{tabular}
  \\[1.5pt]
  \begin{tablenotes}
      \item\label{tnote:a}$^{a}$\citet{Caplan2000}
     \end{tablenotes}
  \caption{Comparison of integrated fluxes (in $10^{-10}\ \rm{erg/s/cm^2}$) between our work and the observations of \citet{Caplan2000}. Coordinates from \citet{Caplan2000} are J1950.0. The SITELLE values for the [\ion{O}{ii}] emission lines are given with and without sigma-clipping.}
  \label{tab:comp_caplan}
\end{table*}

\subsection{Sensitivity}
\label{sec:sens}

We derive the typical sensitivity of the three spectral cubes from the noise maps of each filter. Histograms of the values in the noise maps (not shown here) reveal peaked distributions for each filter, where the typical mean and median values are $\sim$ $1.1\times10^{-16}$, $2.6\times10^{-18}$, and $7.2\times10^{-18}~\rm{erg/s/cm^2}$/\AA\ for the SN1, SN2, and SN3 filters, respectively. The individual emission lines are assumed to be unresolved and their typical widths ($\sigma$ of the $sinc$ function, a factor 3.8 lower than its full width at half maximum) derived from the fitting procedure are $\sim$ 1.3~cm$^{-1}$ for the [\ion{O}{ii}] lines in the SN1 filter, 9.4~cm$^{-1}$ for the H$\beta$ line in the SN2 filter, and 2.9~cm$^{-1}$ for the H$\alpha$ line in the SN3 filter. These widths correspond to effective resolving power in the cubes of about 5500 at the wavelength of the [\ion{O}{ii}] lines, 600 at the wavelength of H$\beta$, and 1400 at the wavelength of H$\alpha$. The resolving power in the SN2 and SN3 filters are those requested, while that in the SN1 filter is significantly lower than the requested 10'000. This is because the SN1 spectral cube was obtained with a power spectrum data reduction method, without any phase correction process, as no phase calibration data were available for that filter (see section \ref{sec:obs}). It is also possible that the [\ion{O}{ii}] are partly resolved: $R=5500$ corresponds to a width of 55~km/s and HH objects can show width of at least a few tens of km/s \citep[see e.g.,][]{Heathcote1998}. These widths, combined with the typical noise amplitude, then lead to flux sensitivities of about $1.5\times10^{-16}$, $2.0\times10^{-17}$, and $2.5\times10^{-17}~\rm{erg/s/cm^2}$ in the SN1, SN2, and SN3 filters, respectively. These values correspond to the sensitivities per SITELLE pixel (0.32\arcsec\ by 0.32\arcsec).

\subsection{Fluxes}
\label{sub:fluxes}

Flux maps are shown in Figure \ref{fig:fluxmaps} for the H$\alpha$, [\ion{O}{iii}]$\lambda$~5007, [\ion{S}{ii}]$\lambda$~6717, and [\ion{N}{ii}]$\lambda$~6584 emission lines. All the flux maps are provided in the appendix. The flux maps reveal the layered structure of the nebula, with the ionized gas emitting from within the H{\sc ii} region and from the illuminated tips of pillar-like structures, while the surrounding molecular clouds appear as dark structures. The observed emission and structure of the region as traced by the various nebular lines covered with the SITELLE data shows a classical H{\sc ii} region, in which the Balmer lines and the highly ionized species trace the inner, diffuse gas, while the [\ion{S}{ii}] and [\ion{N}{ii}] lines are more localized towards the ionization fronts, for instance at the tip of the pillars \citep[e.g., MC15, ][]{weilbacher15}.

Hereafter we compare the fluxes measured in the SITELLE data with those published by \citet{Caplan2000} with large apertures, \citet{GarciaRojas2006} with VLT/UVES, MC15 with VLT/MUSE, and with archival HST/WFC3 narrowband images, to demonstrate the quality of the flux calibration with SITELLE on a target like M~16.

\subsubsection{Large apertures}
\citet{Caplan2000} used a Fabry-Perot spectrometer at the 1.5m telescope in San Pedro M\'artir to measure the integrated fluxes of the [\ion{O}{ii}]$\lambda$3726/3729, H$\beta$, [\ion{O}{iii}]$\lambda$5007, and H$\alpha$ emission lines over several circular apertures of 1\arcmin35\arcsec\ and 3\arcmin11\arcsec\ in diameter. We measure the integrated flux for these emission lines over the same regions in the SITELLE flux maps (see Table \ref{tab:comp_caplan}). When performing the aperture integrations, we first include all pixels within the aperture and then exclude pixels with an amplitude of the $sinc$ function lower than the noise measured for that pixel. The effect is significant, at about 20-30\%, for the fainter [\ion{O}{ii}] lines. For the H$\beta$, [\ion{O}{iii}]$\lambda$5007, and H$\alpha$ lines, however, this clipping has only a very minor effect ($<$1\%) on the integrated fluxes and we thus use the uncertainty maps to compute uncertainties on the integrated fluxes for those lines.

For the [\ion{O}{ii}] lines, we combine all radial velocity components together and report the values with and without sigma-clipping. The overall agreement is good between our measurements and those from \citet{Caplan2000}, though they vary from line to line and from region to region. There seems to be a systematic trend where the agreement is much better for the first two regions than the last two. The five emission lines we measure here are typically within 5 to 15\% of the \citet{Caplan2000} measurements in the first two regions, while they are within 20 to 30\% in the last two regions. The agreement is better overall for the H$\alpha$ line, with a discrepancy lower than 4\% in the first two regions, while it is about 10-15\% for the H$\beta$ and [\ion{O}{iii}]$\lambda$5007 emission lines. For the [\ion{O}{ii}] emission lines, because the measurements are more uncertain, it is more difficult to assess the agreement with \citet{Caplan2000}. Our measurements for these two lines are within a few \% to 30\% of the \citet{Caplan2000} values depending on the region and whether or not we apply a sigma-clipping to the flux images.

\begin{table*}[h]
\centering
  \begin{tabular}{c c | c c c | c c}
    \hline
    \hline
    \multicolumn{2}{c|}{Line} & \multicolumn{3}{c|}{Flux ratio} & \multicolumn{2}{c}{Flux} \\
	  & & $^a$ & SITELLE & MUSE & SITELLE & MUSE \\
    \hline
	\multirow{ 5}{*}{SN3} & [\ion{N}{II}]$\lambda$6548  & 104.213 &  114$\pm$12 & 107$\pm$11 & 42$\pm$4 & 38$\pm$3 \\
	& ~[\ion{N}{II}]$\lambda$6584  & 326.951 & 306$\pm$30 & 330$\pm$32 & 113$\pm$10 & 115$\pm$9 \\
	& H$\alpha$                    & 614.888 &  643$\pm$3 &  679$\pm$4 & 239$\pm$11 & 240$\pm$14 \\
	& ~[\ion{S}{II}]$\lambda$6717  & 51.101  &   49$\pm$7 &   51$\pm$7 &   18$\pm$2 &  18$\pm$2 \\
	& ~[\ion{S}{II}]$\lambda$6731  & 65.429  &   57$\pm$8 &   61$\pm$9 &   21$\pm$3 &  22$\pm$2 \\
	\hline
    \multirow{ 3}{*}{SN2} & H$\beta$ & 100   &        100 &        100 &   37$\pm$2 & 36$\pm$2 \\
	& ~[\ion{O}{III}]$\lambda$4960 & 29.222  &   31$\pm$3 &   36$\pm$2 &   12$\pm$2 & 13$\pm$1 \\
	& ~[\ion{O}{III}]$\lambda$5007 & 88.392  &  100$\pm$8 &  111$\pm$8 &   38$\pm$4 & 39$\pm$4 \\
	\hline
    \multirow{ 8}{*}{SN1} & ~[\ion{O}{II}]$\lambda$3726  & 88.191  &   46$\pm$2 & N/A & & \\
    & main                         &         &            &       & 9.3$\pm$0.3 & N/A \\
	& red                          &         &            &       & 6.4$\pm$0.4 & N/A \\
	& blue                         &         &            &       & 1.4$\pm$0.3 & N/A \\
	& ~[\ion{O}{II}]$\lambda$3729  & 69.245  &   40$\pm$1 &   N/A & \\
    & main                         &         &            &       & 8.9$\pm$0.3 & N/A \\
	& red                          &         &            &       & 4.6$\pm$0.5 & N/A \\
	& blue                         &         &            &       & 1.3$\pm$0.1 & N/A \\
	\hline
  \end{tabular}
    \\[1.5pt]
  \begin{tablenotes}
      \item\label{tnote:a}$^{a}$\citet{GarciaRojas2006}
     \end{tablenotes}
  \caption{Comparison of observed integrated fluxes and flux ratios between our work, \citet{GarciaRojas2006}, and MC15. Fluxes are in units of $10^{-14}\ \rm{erg/s/cm^2}$.}
  \label{tab:comp_GR06}
\end{table*}

\subsubsection{VLT/UVES}

\citet{GarciaRojas2006} used VLT UVES to measure faint emission lines in M~16. These authors extracted the flux in a 3\arcsec-by-8.5\arcsec\ region centered 48\arcsec\ north and 40\arcsec\ west of BD-13~4930. We measure the integrated fluxes for the emission lines detected in the SITELLE data over the same region and normalized them to H$\beta$ = 100 to directly compare them to the observed UVES values (see Table \ref{tab:comp_GR06}). Because there might be a slight mismatch in the astrometry, we scatter the virtual slit on the SITELLE image within a region of $\pm3$\arcsec\ and provide the corresponding range of values in the Table.

For all emission lines in the SN2 and SN3 filters, the agreement with the values of \citet{GarciaRojas2006} is very good. The typical discrepancy on the H$\beta$ normalized values is between 5 and 10\%, except for the [\ion{S}{II}]$\lambda$6731 line which is about 12\% lower in the SITELLE data. The [\ion{O}{II}] to H$\beta$ line ratios are significantly lower in the SITELLE data than in the VLT/UVES observations with almost a factor 2 discrepancy between the two sets of measurements.

We note, however, that the "line intensities were measured integrating all the flux in the line between two given limits and over a local continuum estimated by eye", according to \citet{GarciaRojas2006}, which means the uncertainties on their line ratios might thus be large, especially for the faintest lines.

\subsubsection{VLT/MUSE}
MC15 obtained nine spectral cubes on the Pillars of Creation in M~16 with MUSE, covering a total field of view of about 3\arcmin\ by 3\arcmin\ with a spectral range from 4750 to 9350~\AA. We perform a pixel-to-pixel comparison of the SITELLE line fluxes derived from the $sinc$ fit and the MUSE line fluxes derived from Gaussian fits, by reprojecting the SITELLE flux maps onto the MUSE pixel grid. Because SITELLE extends farther in the blue than MUSE, only the lines in the SN3 and SN2 filters are compared. In addition, we correct for a slight astrometric offset between the two datasets (about five MUSE pixels or 1\arcsec).

\begin{figure*}[t]
\centering
\includegraphics[width=0.33\linewidth]{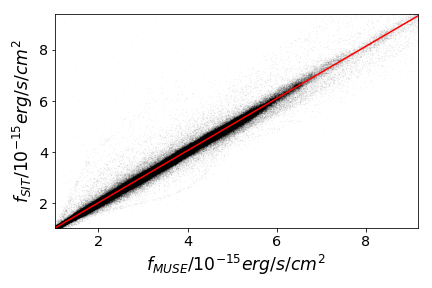}
\includegraphics[width=0.33\linewidth]{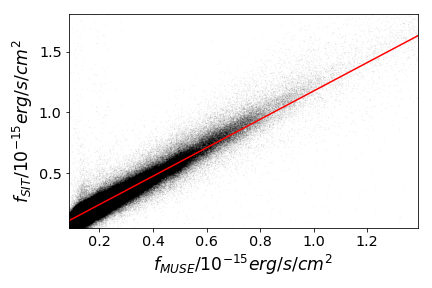}
\includegraphics[width=0.33\linewidth]{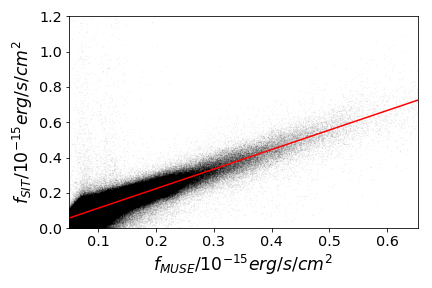}
\caption{Pixel-to-pixel comparison between the line fluxes measured in MUSE and SITELLE data for H$\alpha$, [\ion{N}{ii}]$\lambda$6548, and [\ion{S}{ii}]$\lambda$6717 (left to right). The red lines show linear regressions to the data points.}
\label{fig:sit_v_muse}
\end{figure*}

The correlation plots and linear fits to these plots show an excellent agreement between the flux maps from MUSE and SITELLE (see examples in Figure \ref{fig:sit_v_muse}). Where the sensitivity advantage of MUSE over SITELLE becomes visible (e.g., for the [\ion{S}{ii}] emission line), the scatter becomes larger, though the correlation remains excellent. Table \ref{tab:sit_v_muse} lists the Spearman's rank correlations coefficients, which are above 0.97 for the majority of the maps. However, for the [\ion{S}{ii}] emission lines, the faintest in this comparative analysis, the correlation coefficient is only 0.72. The table also gives the slopes derived from a linear fit with three $3\sigma$ clipping iterations, along with the median of the ratios between the SITELLE and MUSE line fluxes. The slopes are typically within 5 to 10\% of unity, except for the [\ion{S}{ii}] doublet and [\ion{N}{ii}]$\lambda$6548 emission lines. The median of the ratios are, with the same exceptions, within 5 to 10\% of unity. The correlations are thus excellent overall for both the SN3 and SN2 filters. We note that the slopes and median values are more likely greater than 1, especially for the faint emission lines, which might be due to the difference in sensitivity between the two data sets. We also compute the MUSE integrated fluxes over multiple regions including those observed by \citet{Caplan2000} and \citet{GarciaRojas2006}, using an identical method to that used for SITELLE. The agreement between MUSE and SITELLE on the integrated fluxes is again confirmed at a lower than 5 to 10\% level for most emission lines (see Tables \ref{tab:comp_caplan} and \ref{tab:comp_GR06}).

The agreement in terms of photometric calibration is thus remarkable between two of the most recent ground-based, optical, integral field spectrographs, despite their very different designs, calibration procedures, and reduction pipelines. This comparison between MUSE and SITELLE additionally highlights the expected differences in terms of sensitivity reached by both instruments: the VLT instrument, on a 8-m class telescope, covers the entire visible spectral range at once and detects significantly fainter emission than SITELLE at CFHT. However, SITELLE, on a 4-m class telescope, can access a wider field of view, at bluer wavelengths, and provide an adjustable spectral resolution.

\begin{table}[t]
\centering
  \begin{tabular}{c c c c c}
    \hline
    \hline
    & Line & $\rho$ & Slope & Median \\
    \hline  
    \multirow{ 3}{*}{SN2}& H$\beta$ & 0.98 & 1.08 & 1.08 \\
    &~[\ion{O}{III}]$\lambda$4960 & 0.97 & 0.93 & 0.94 \\
    &~[\ion{O}{III}]$\lambda$5007 & 0.99 & 0.96 & 0.96 \\
    \hline
    \multirow{ 5}{*}{SN3}& ~[\ion{N}{II}]$\lambda$6584  & 0.98 & 1.04 & 1.03 \\
    &H$\alpha$                    & 0.99 & 1.02 & 1.02 \\
    &~[\ion{N}{II}]$\lambda$6548  & 0.90 & 1.17 & 1.23 \\
	&~[\ion{S}{II}]$\lambda$6717  & 0.78 & 1.11 & 1.15 \\
    &~[\ion{S}{II}]$\lambda$6731  & 0.72 & 1.13 & 1.15 \\
    \hline
  \end{tabular}
  \caption{Results of the pixel-to-pixel comparison between the line fluxes measured in MUSE and SITELLE data for all emission lines covered by both instruments.}
  \label{tab:sit_v_muse}
\end{table}

\begin{figure*}[!t]
\centering
\includegraphics[width=0.33\linewidth]{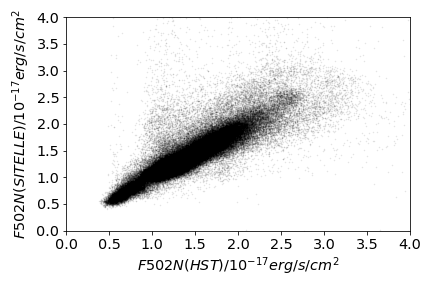}
\includegraphics[width=0.33\linewidth]{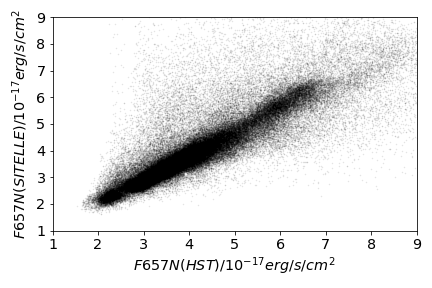}
\includegraphics[width=0.33\linewidth]{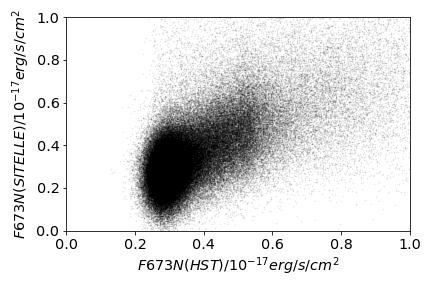}
\caption{Pixel-to-pixel comparisons between the fluxes measured in HST/WFC3 F502N, F657N, and F673N filters (left to right) and their equivalent in the SITELLE data.}
\label{fig:sit_v_hst}
\end{figure*}

\begin{table*}[!t]
\centering
  \begin{tabular}{c c c c c c c}
    \hline
    \hline
    & & Aperture \#1 & Aperture \#2 & Aperture \#3 & Aperture \#4 \\
	\hline
	\multicolumn{2}{c}{RA (J2000)}  &  18 18 51.03 &  18 18 48.91 &  18 18 50.06 &  18 18 48.74 \\ 
	\multicolumn{2}{c}{DEC (J2000)} & -13 49 04.01 & -13 49 04.94 & -13 49 33.68 & -13 49 49.68 \\ 
	\hline
	\multirow{5}{*}{F502N} & HST/WFC3    & 2.56 & 2.99 & 3.04 & 3.29 \\
	                       & SIT         & 2.43 & 2.70 & 2.89 & 3.25 \\
	                       & SIT$_{s}$   & 2.34 & 2.60 & 2.78 & 3.10 \\ 
	                       & SIT$_{g}$   & 2.33 & 2.59 & 2.77 & 3.10 \\ 
                           & MUSE        & 2.72 & 2.75 & 3.09 & 3.39 \\ 
	\hline
	\multirow{5}{*}{F657N} & HST/WFC3    & 8.39 & 7.92 & 7.57 & 11.6 \\ 
	                       & SIT         & 7.95 & 6.85 & 7.07 & 10.8 \\
	                       & SIT$_{s}$   & 7.83 & 6.76 & 6.94 & 10.7 \\ 
	                       & SIT$_{g}$   & 7.83 & 6.76 & 6.94 & 10.7 \\ 
                           & MUSE        & 8.08 & 6.42 & 6.87 & 10.3 \\
	\hline
   	\multirow{5}{*}{F673N} & HST/WFC3    & 0.86 & 0.61 & 0.58 & 1.22 \\ 
	                       & SIT         & 0.81 & 0.52 & 0.54 & 1.04 \\
	                       & SIT$_{s}$   & 0.90 & 0.82 & 0.69 & 1.37 \\ 
	                       & SIT$_{g}$   & 0.89 & 0.87 & 0.73 & 1.39 \\ 
                           & MUSE        & 0.96 & 0.61 & 0.61 & 1.18 \\
	\hline
  \end{tabular}
  \caption{Results of the comparison between the narrow band images from HST/WFC3, SITELLE, and MUSE. For SITELLE, we use (1) the original data, (2) a thinly sampled "sinc" function identical to the fit, and (3) a thinly sampled Gaussian model with identical integrated flux, width, and position as the "sinc" fit. All values are in $10^{-14}~\rm{erg/s/cm^{-2}/A}$.}
  \label{tab:sit_v_hst}
\end{table*}

\subsubsection{HST/WFC3}

We finally compare the SITELLE data to narrow-band imaging obtained as part of the Hubble Heritage program\footnote{\url{http://heritage.stsci.edu/2015/01/index.html}} with HST/WFC3 and the F502N, F657N, and F673N filters. We compute the fluxes that would be detected by HST/WFC3 if the emitted spectrum was that from the SITELLE spectral cubes, for each pixel in the field of view. We use the transmission curve of the filters\footnote{See for instance \url{http://svo2.cab.inta-csic.es/svo/theory/fps3/index.php?id=HST/WFC3_UVIS1.F502N&&mode=browse&gname=HST&gname2=WFC3_UVIS1#filter}} and the flux definition given by Equation \ref{eq:hst}:

\begin{equation}
	f = \frac{\int F_\lambda \tau_\lambda \lambda d\lambda}{\int \tau_\lambda \lambda d\lambda}
\label{eq:hst}
\end{equation}

where $F_\lambda$ is the flux density of the target given by the SITELLE data and $\tau_\lambda$ is the transmission curve of the filter.

The pixel-to-pixel correlation between the SITELLE and the HST data (see Figure \ref{fig:sit_v_hst}) is excellent in the F502N and F657N filters. Because the emission lines in the F673N filters are fainter, the pixel-to-pixel correlation is not as good, with a significant scatter of the SITELLE measurements below $\simeq5\times10^{-18}~\rm{erg/s/cm^{-2}}$.

\begin{figure*}[!t]
\centering
\includegraphics[width=0.24\linewidth]{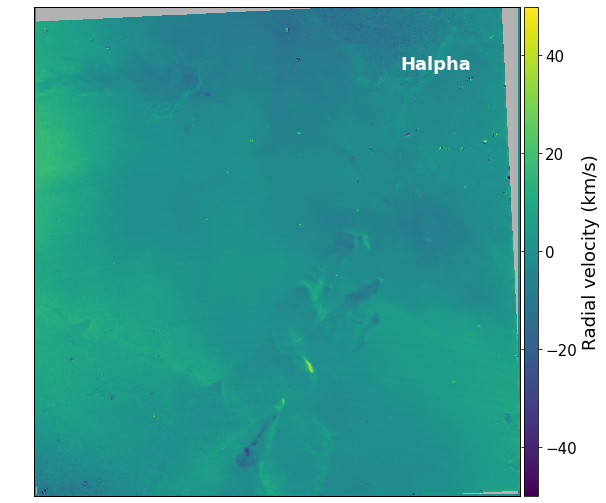}
\includegraphics[width=0.24\linewidth]{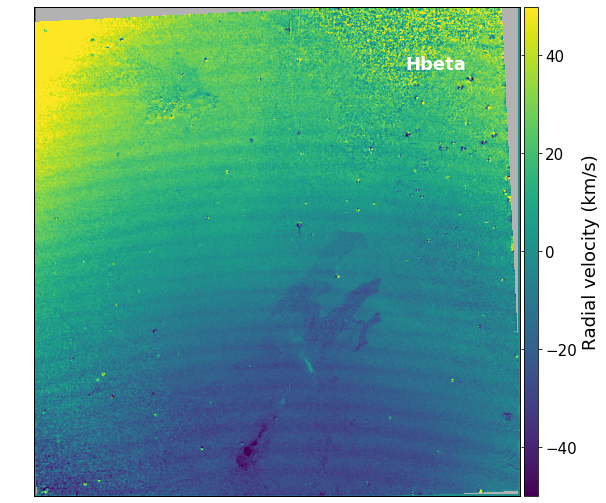}
\includegraphics[width=0.24\linewidth]{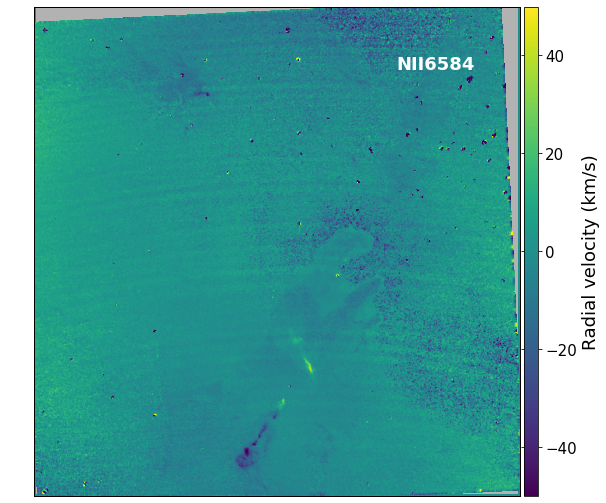}
\includegraphics[width=0.24\linewidth]{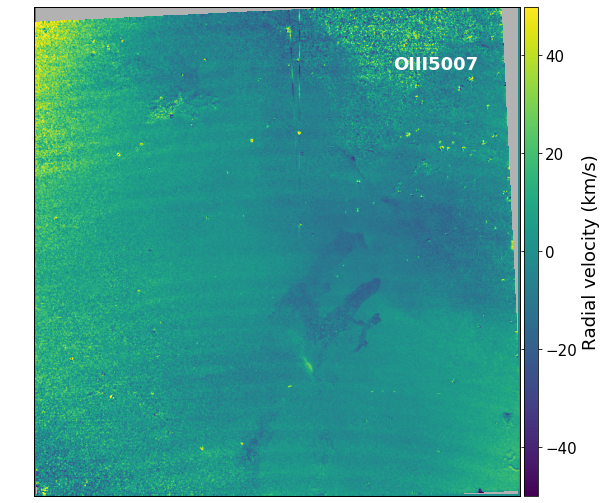}\\
\includegraphics[width=0.24\linewidth]{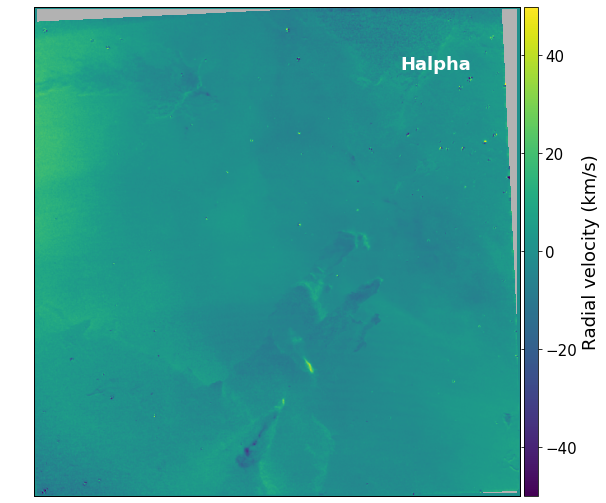}
\includegraphics[width=0.24\linewidth]{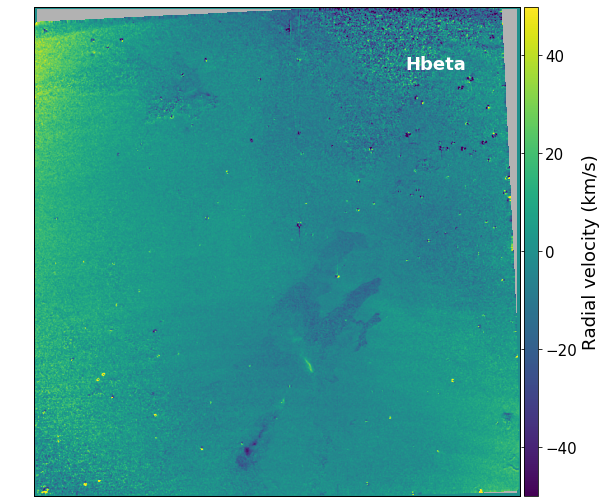}
\includegraphics[width=0.24\linewidth]{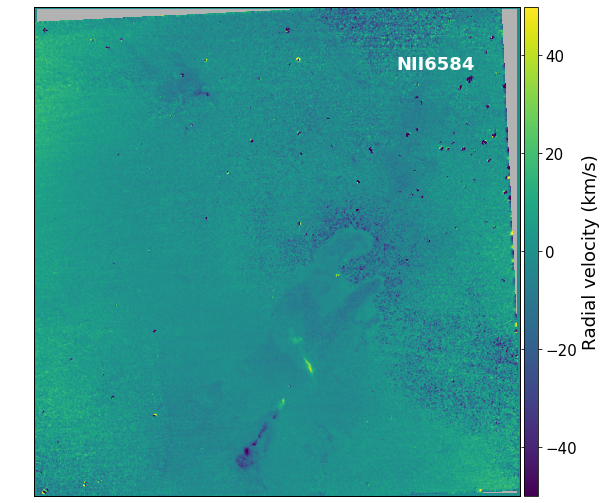}
\includegraphics[width=0.24\linewidth]{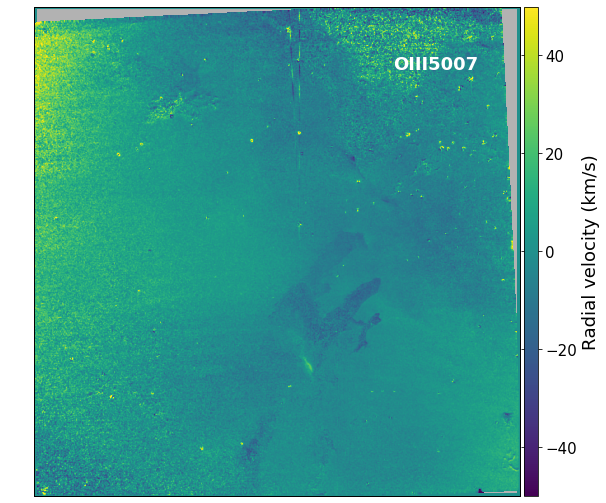}
\caption{Emission line radial velocity maps for some of the emission lines that clearly show velocity structure. From left to right: H$\alpha$, H$\beta$, [\ion{N}{ii}]$\lambda$6584, and [\ion{O}{iii}]$\lambda$5007. The top row is before correction but after subtraction of the median value, the bottom row is after correction (see text for details). All maps are smoothed with a 2-pixel wide Gaussian and the scale ranges from -50~km/s to +50~km/s.}
\label{fig:velmaps_corr}
\end{figure*}

We then compute the integrated SITELLE and HST fluxes in the HST filters within several circular apertures of 8\arcsec\ radius (see Table \ref{tab:sit_v_hst}). For the F502N filter, which mostly traces the [\ion{O}{iii}]~5007 emission line, the agreement is excellent, with a discrepancy that ranges from 1 to 10\% between the HST and SITELLE data. For the F657N filter, which mostly traces the H$\alpha$ emission line, the agreement is very similar to that in the F502N filter, with a discrepancy between the HST and SITELLE data that ranges from 5 to 16\%. Similarly, the discrepancy between the HST and SITELLE data for the F673N filter, which mostly traces the [\ion{S}{ii}] doublet, ranges from 5 to 17\%. We note however that the SITELLE integrated fluxes are systematically lower than those in the HST data. We perform the same measurements using the MUSE data and find they are in a slightly better agreement with the HST values than the SITELLE data. The discrepancy between MUSE and HST is randomly distributed within 9\% for the F502N filter, within 11\% for the F673N filter, while MUSE leads to systematically lower values than HST by 4 to 23\% for the F657N filter.

We then estimate the impact of the emission lines $sinc$ profile on the comparison. Because the HST filters are narrow, the wings of some $sinc$ lines spread outside the filter from within while other lines spread their wings inside from without. We also investigate whether the spectral sampling of the SITELLE data affects the measurements presented in this section. To assess the significance of these effects, we first use oversampled $sinc$ profiles to model the emission lines instead of the data and compute the integrated flux within the HST/WFC3 filters the same way as above. The $sinc$ profiles have the widths, amplitudes, and positions derived in the fit but they are sampled on a wavenumber grid about three times thinner than the real data. We also use oversampled gaussian profiles instead of the $sinc$ profiles, with the same widths and positions, but with amplitudes multiplied by a factor $\sqrt(\pi/2)$ to maintain the same integrated fluxes. The resulting measurements within the HST filters are indicated as SIT$_{s}$ and SIT$_{g}$ in Table \ref{tab:sit_v_hst}. These measurements are very similar to those derived directly from the data, though they seem to be systematically lower, except within the F673N filter, where they are significantly larger. We therefore argue that the sampling and line profile of the SITELLE data may affect narrow filter flux measurement comparisons, although it seems insignificant in most cases.

\subsubsection{Summary of the photometric comparisons}

Our conclusions on the photometric calibration of SITELLE are that the relative and absolute flux measurements are typically within 10\% of previously published measurements. Some comparisons lead to larger discrepancies \citep[the SN1 comparison with][]{GarciaRojas2006} but can be explained by a large uncertainty in the continuum level in the previously published results. Other discrepancies can typically be justified by the difference in sensitivity of the various instruments. To mitigate this effect, one could bin the spectral cubes of SITELLE before modeling the emission lines for an increased S/N at a cost of a decreased image quality.

\subsection{Radial velocities}
\label{sec:vel}

Several radial velocity maps are shown in Figure \ref{fig:velmaps_corr}. These maps were derived from the best fit centroid positions of each emission line, in $\rm{cm^{-1}}$, converted into a radial velocity according to the reference wavelengths listed in Table \ref{ref_wave}.

\begin{figure*}[!t]
\centering
\includegraphics[scale=0.9,trim=2cm 1cm 0cm 0cm]{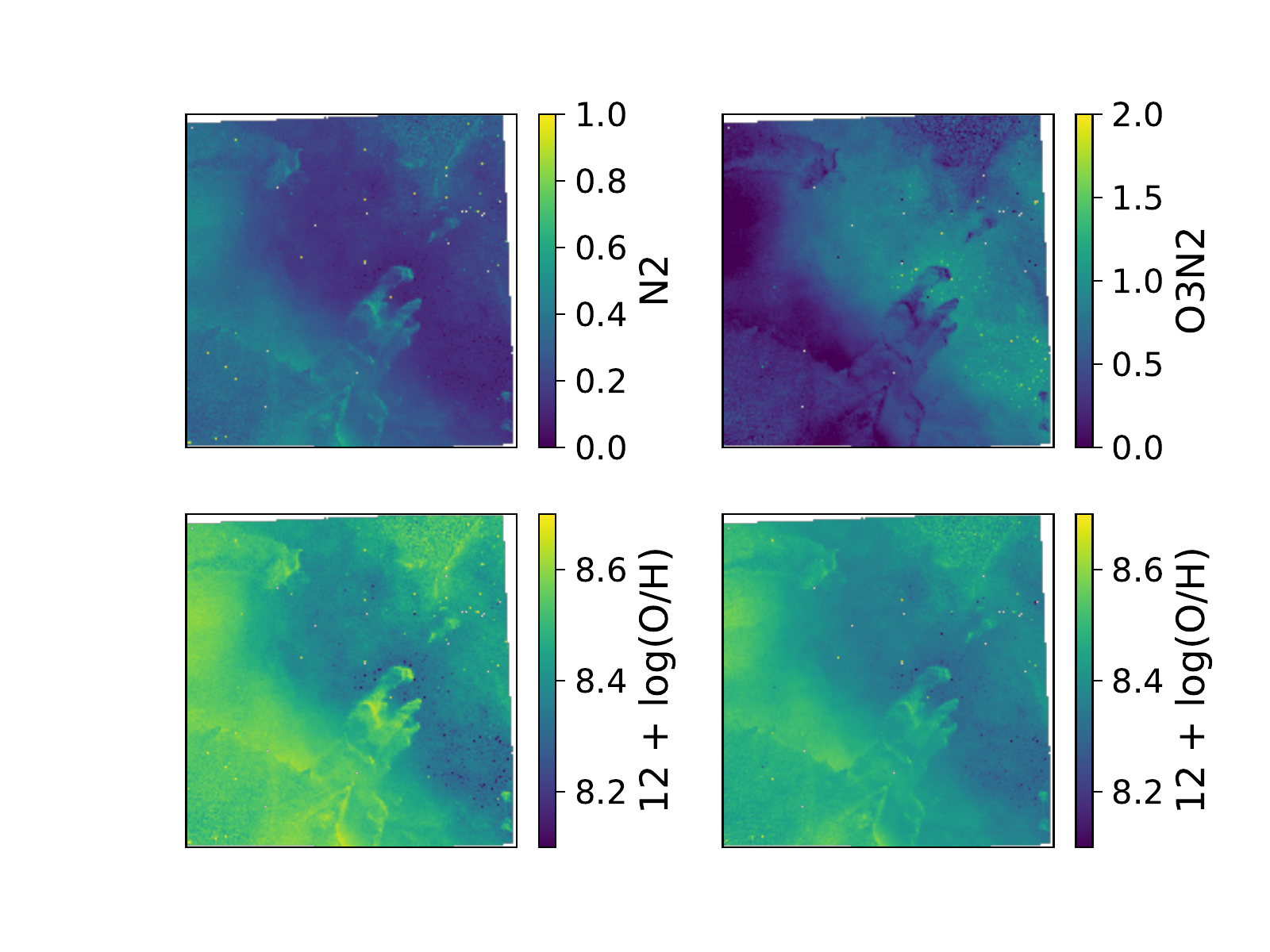}
\caption{Line ratio maps (upper left is N2 and upper right is O3N2), as well as the derived oxygen abundance from the two (lower left and lower right, respectively). The lower panels are scaled to the same limits to illustrate the different values obtained from the two tracers.}
\label{ox}
\end{figure*}

While some features are noticeable in the nebula (e.g., the Pillars of Creation and HH~216), large scale arcs and gradients are also apparent, due to an imperfect wavelength calibration of the SITELLE data. This is the consequence of using external calibration data to infer the phase corrections that are then applied to the science cubes, rather than estimating the phase correction internally from the science cubes themselves (see section~\ref{sec:obs}). Indeed, these calibration data are not necessarily obtained in the exact same conditions as the science cubes. The resulting errors are mostly linked to variations of the effective incidence angle of the light on the detectors, which explain their arc shape. To remove those instrumental defects, we use a laser calibration map that is part of the calibration data and provides a wavelength for each pixel in the field of view. With this map, we identify the arc pattern with a width of 0.01~nm, and for each emission line radial velocity map, we then compute and subtract the median value within each arc. We therefore also remove the absolute median radial velocity towards M~16 but highlight the small scale variations. Examples of corrected radial velocity maps, smoothed with a Gaussian kernel ($\sigma=2$ pixels) are shown in Figure~\ref{fig:velmaps_corr}. The arc and gradient features have almost entirely been removed and the agreement between emission lines coming from different filters is excellent. A new version of the SITELLE data reduction pipeline is currently being implemented to take care of these artifacts and future papers will not show these anymore.

The most striking feature in the radial velocity maps for the SN2 and SN3 emission lines is related to HH~216: a jet comprising multiple blobs at negative relative radial velocities of at least 30~km/s and a bow shock feature on the opposite side at positive relative radial velocities of at least 30~km/s, relative to the motion of the entire region. In the higher spectral resolution SN1 cube, multiple radial velocity components of the [\ion{O}{ii}] doublet can be identified (see Figure \ref{fig:spectra}). The main component is that related to the whole nebula. In addition, a blueshifted component is detected towards the clumps of HH~216 and a redshifted component is detected towards the bow shock feature of HH~216. The shift in radial velocity brings the red component of the [\ion{O}{ii}]$\lambda$3726 emission line at a wavelength very similar to that of the blue component of the [\ion{O}{ii}]$\lambda$3729 emission line. As a consequence, the flux maps for these two specific components is deemed unreliable. However, the blueshifted component of the [\ion{O}{ii}]$\lambda$3726 emission line and the redshifted component of the [\ion{O}{ii}]$\lambda$3729 emission line are clearly exhibiting a radial velocity shift of about 70~km/s. In the SN2 and SN3 cubes, which are at spectral resolutions of 1400 at most, only a single radial velocity component can be identified and the radial velocity maps simply reveal a shift in the average radial velocity along the line of sight. These maps, however, correspond to brighter emission lines and thus to higher S/N. We thus use both in section \ref{sec:hh} where we discuss HH~216 in more detail.

\section{Line ratio diagnostics}
\label{sec:diag}

Ratios of nebular lines are powerful diagnostics to probe physical quantities such as the electron density, the electron temperature, elemental abundances, the degree of ionization, and the contribution of shock- and photo-ionization. Large-scale electron density and temperature maps of the central pillars have been reported in MC15, and given the higher resolving power of MUSE in the 6000-7000 {\AA} regime (in which the sensitive [\ion{S}{ii}] and [\ion{N}{ii}] lines are found), we do not perform the same density and temperature analysis here. 

There are various methods for deriving gas-phase metallicities. The classical $T_{e}$-{\it method} uses temperature-sensitive line ratios to determine the electron temperature $T_{\mathrm{e}}$, that is ([\ion{N}{ii}]$\lambda$6548 + [\ion{N}{ii}]$\lambda$6584) / [\ion{N}{ii}]$\lambda$5755 \citep{osterbrock}, from which ionic abundances can then be derived, followed by the derivation of the total elemental abundance with an appropriate ionization correction factor (e.g., \citealt{perez17}). Despite being widely used to derive abundances in Galactic and extragalactic H{\sc ii} regions (e.g., \citealt{lin17}, \citealt{perez17}, \citealt{pilyugin16}, \citealt{orion}, \citealt{westmoquette13}, \citealt{monreal12}, \citealt{bresolin12}), it comes with the caveat that auroral lines (needed to determine T$_{e}$) are typically very faint, thus requiring high S/N observations. Another much used method to determine gas-phase metallicities is the so-called \textit{strong-line method}, which is based on abundance-tracing emission line ratios. Together with empirical calibrations (for example abundances derived from the $T_{e}$-{\it method}), the intensity of strong emission lines is therefore used as an indirect abundance measure. With the wavelength coverage of SITELLE, we exploit the commonly used line ratios, O3N2 and N2 (e.g., \citealt{monreal11}) in combination with the empirical relations derived in \citet{marino13}, to derive the oxygen abundance of the observed region (Figure~\ref{ox}). The O3N2 and N2 line ratios are defined as follows:

\begin{equation}
\mathrm{O3N2 = log\bigg(\frac{[\ion{O}{iii}]\lambda5007}{H\beta}\times\frac{H\alpha}{[\ion{N}{ii}]\lambda6583}\bigg)}
\label{o3n2}
\end{equation}

\begin{equation}
\mathrm{N2 = log\bigg(\frac{[\ion{N}{ii}]\lambda6583}{H\alpha}\bigg)}
\label{n2}
\end{equation}

\begin{figure}
\includegraphics[scale=0.45]{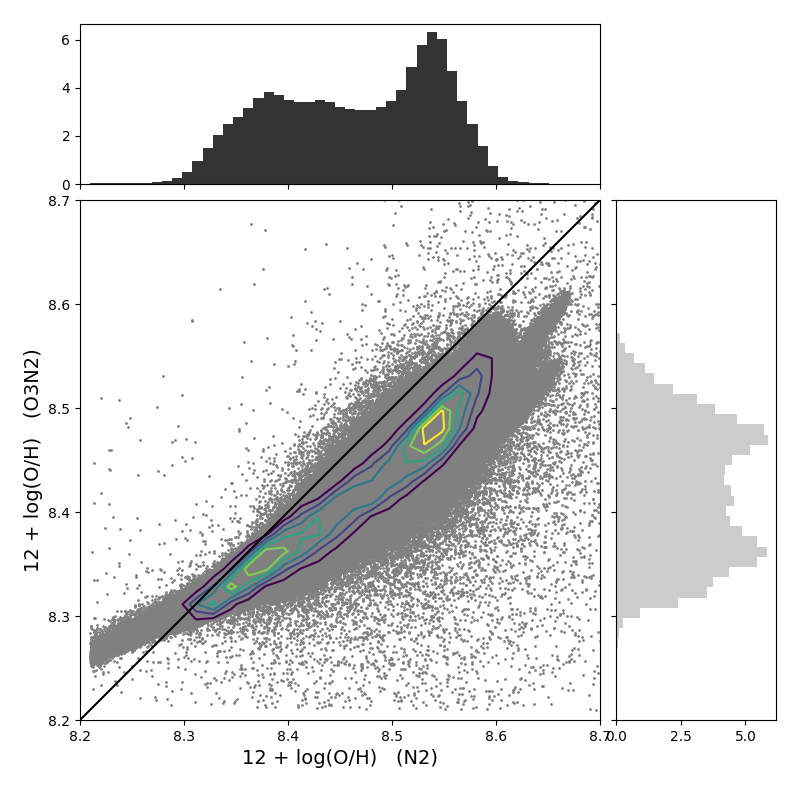}
\caption{Oxygen abundances derived from the two line ratios, N2 and O3N2. The contours trace the highest concentration of data points, with levels at 20$\cdot10^{3}$, 40$\cdot10^{3}$, 60$\cdot10^{3}$, 80$\cdot10^{3}$, 100$\cdot10^{3}$, and 120$\cdot10^{3}$. The histograms are normalized to peak, and in black is shown the one-to-one line.}
\label{oxi_comp}
\end{figure}

\begin{figure}
\includegraphics[scale=0.6]{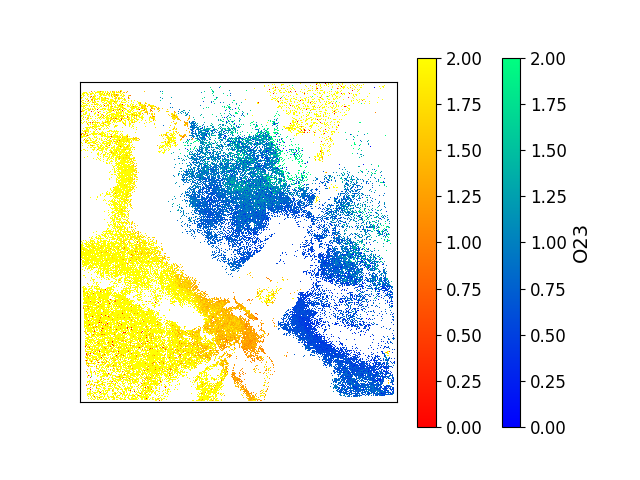}
\caption{Masked O23 map only showing the pixels in the two peaks of the selected contour level from Figure~\protect\ref{oxi_comp}(b). The red and blue colormaps trace pixels from the higher and lower peaks, respectively.}
\label{metal_mask}
\end{figure}

\begin{figure}
\includegraphics[scale=0.55]{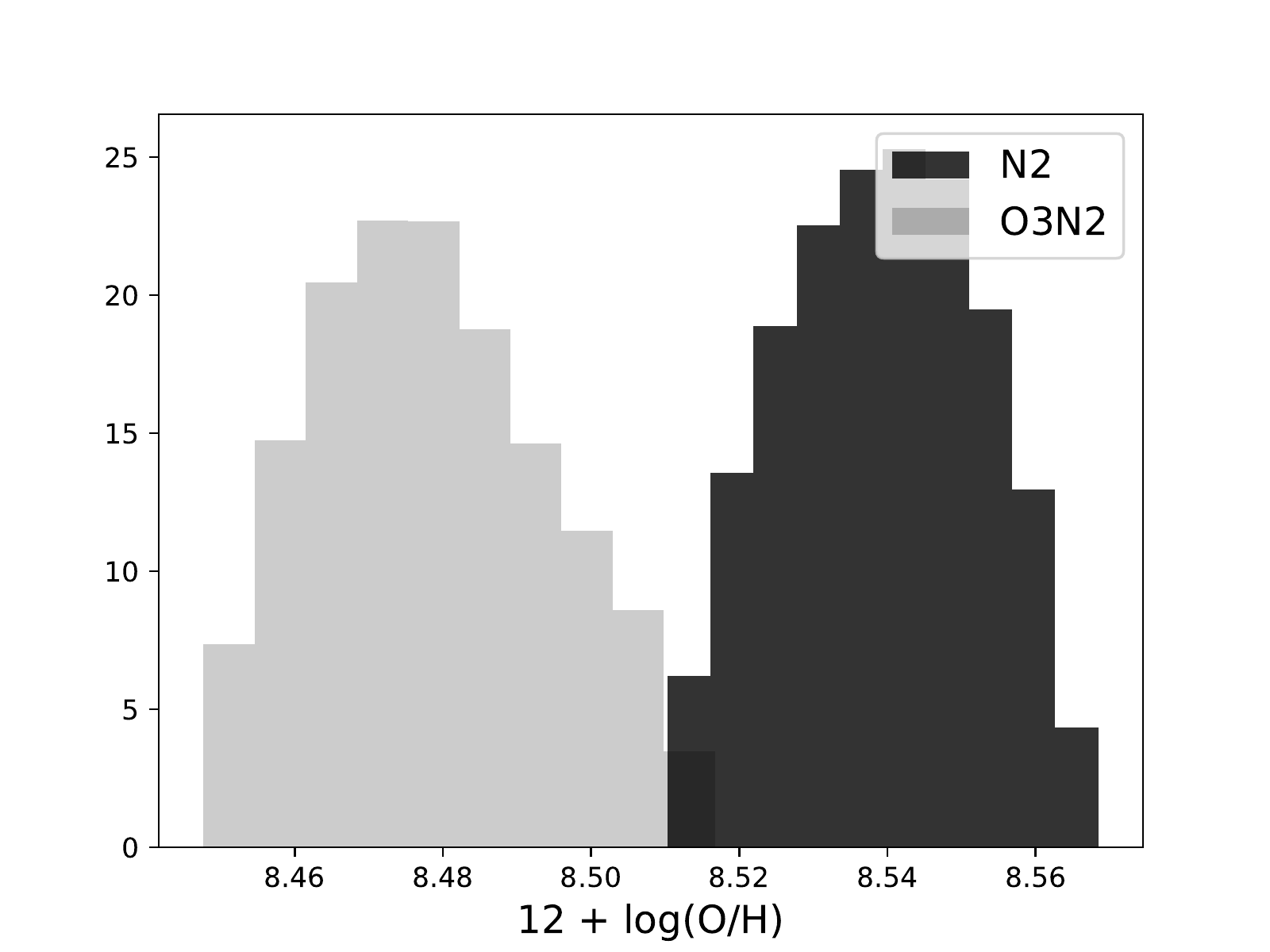}
\caption{Histograms of the data points in the higher-valued peak in Figure~\protect\ref{oxi_comp}(b) for the metallicity derived from N2 and O3N2.}
\label{metal_ion_hist}
\end{figure}

O3N2 is a ratio of the strongest observed emission lines and, given that we de-redden the line maps according to the Balmer decrement (MC15), the O3N2 ratio is effectively tracing [\ion{O}{iii}]/[\ion{N}{ii}], hence a ratio of collisionally-excited lines. N2 on the other hand suffers less from the reddening correction than O3N2 (or [\ion{O}{iii}]/[\ion{N}{ii}]), given that it is a ratio of two lines separated by only a few tens of {\AA} \citep{marino13}. As shown in Figures \ref{ox} and \ref{oxi_comp}, when compared to N2, the O3N2 diagnostic ratio yields O/H values which are lower for regions with greater density and temperature variations, that is structures such as the tips of the pillars, and a clear bimodal trend is seen for both tracers (Figure~\ref{oxi_comp}).

We note that computing maps of emission line ratios for spatially resolved regions such as this data set (in which single structures rather than integrated values over entire regions/galaxies are seen) comes with a series of caveats (see, for example, the discussion in \citealt{ercolano12}). Indeed, the variations of the oxygen abundance and similar derived values across spatially resolved structures is likely due to local variations of physical parameters such as the density and the temperature (see e.g., \citealt{orion,mcleod19}) which trace the ionization structure of the nebula, leading to regions with a higher degree of ionization showing lower O/H values.

\begin{table}[t]
    \centering
    \caption{Mean oxygen abundance values (denoted by $\overline{X}$) obtained from the two line ratios N2 and O3N2 from all data points, and after correcting for ionization.}
    \begin{tabular}{c c c c c }
         \hline
         \hline
         & N2$_{all}$ & N2$_{ion}$ & O3N2$_{all}$ & O3N2$_{ion}$ \\
         \hline
         $\overline{X}$ & 8.47 $\pm$ 0.08 & 8.54 $\pm$ 0.06 & 8.42 $\pm$ 0.01 & 8.48 $\pm$ 0.01 \\
         \hline
         \hline
    \end{tabular}
    \label{abun}
\end{table}

\begin{figure*}
\centering
\includegraphics[scale=0.5, trim= 0cm 0cm 4cm 0cm]{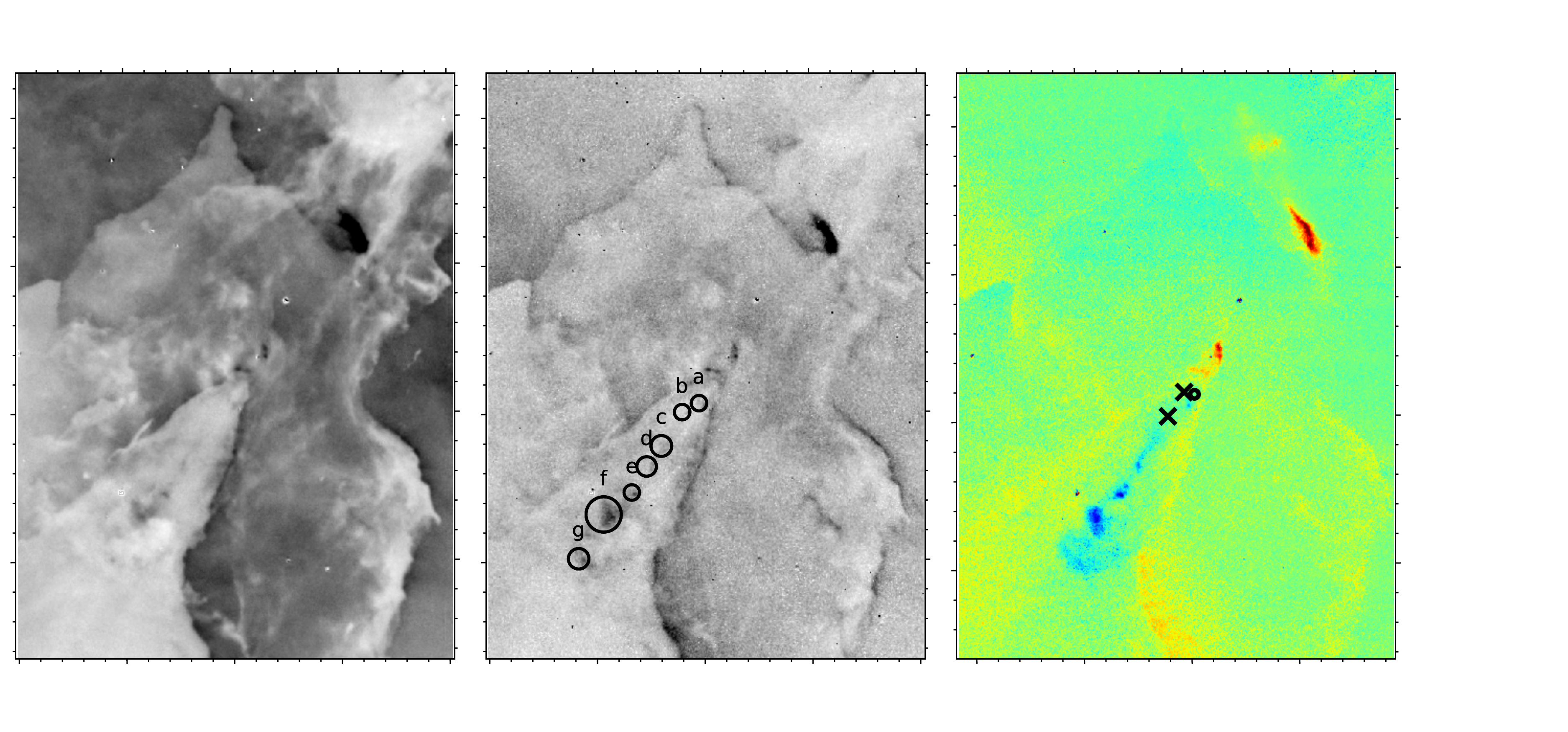}
\caption{Left and middle panels are H$\alpha$ and [\ion{S}{ii}]$\lambda$6717 flux maps, respectively, while the right panel shows the H$\alpha$ radial velocity map. The radial velocity map shows radial velocities, relative to the mean value of the surrounding matter, and is scaled from -45 to 40 km s$^{-1}$. The circles labelled {\it a} through {\it g} identify the seven blueshifted knots. The upper and lower crosses in the right panel indicate the position of HH-N and HH-S respectively, while the black circle indicates the position of a point source identified in the NIR data (see Figure~\ref{nir}).}
\label{hh216}
\end{figure*}

To visualize the dependence of the derived O/H values on the ionization structure of the nebula and demonstrate that low O/H values correspond to higher degree of ionization, we exploit the line ratio O23 $\equiv$ [\ion{O}{ii}]/[\ion{O}{iii}], a tracer of the degree of ionization. This is shown in Figure~\ref{metal_mask}, where we plot the data points within the fourth contour level of Figure~\ref{oxi_comp}, which encompasses two distinct peaks, one at lower and one at higher O/H values. From this contour level we extract the data points corresponding to the two local maxima and trace these back to their spatial correspondence in the O23 map. This shows that the data points with lower O/H values, which coincide with the peak at lower values in the bimodal histograms in Figure~\ref{oxi_comp}, correspond to the lower O23 values (higher ionization, blue color scale) found in the H{\sc ii} region material, while the higher O/H values correspond to less ionized (higher O23 values) regions (red color scale). We therefore select the higher contour peak (red data points in Figure~\ref{metal_mask}) as representative of the gas-phase metallicity. Figure~\ref{metal_ion_hist} shows the resulting histograms of O/H  derived from N2 and O3N2 for the data points selected as just described. Clearly, there is still a systematic discrepancy between the two, O3N2 leading to lower O/H values, which is likely due to the combination of O3N2 being a ratio of two lines with different ionization energies, and the two lines being very far in terms of wavelength. For both line ratios, when comparing the mean values of O/H found by including all data points and those found by selecting only the data points in the higher contour peak, we find that that the O/H values are lower when considering all data points (see Table \ref{abun}), and by taking the mean of the N2- and O3N2-derived O/H values, we find $\overline{X}=8.51\pm 0.01$, which is in good agreement with $\overline{X}_{\mathrm{R10}}=8.52\pm 0.03$ \citep{rodriguez10}. 

We therefore conclude that when deriving gas phase metallicities for spatially resolved H{\sc ii} regions via the \textit{strong-line method} rather than via the direct $T_{e}$-{\it method}, variations of the density, temperature, and degree of ionization result in non trivial abundance variations throughout the region. Multiple line ratios tracing different ionization states (e.g., [\ion{O}{ii}] and [\ion{O}{iii}]) are needed in order to account for the dependency on the degree of ionization on spatially resolved scales.

\section{A multi-wavelength analysis of HH 216}
\label{sec:hh}

The HH 216 flow was first discovered by \citet{Meaburn1982}, who named it M16 HH1, on [\ion{S}{ii}]$\lambda6717,31$ narrow filter plate photographs with the 1.2-m SRC Schmidt camera. In combination with molecular millimeter observations, \cite{andersen04} used VLT near-infrared $J_{\mathrm{s}}$, $H$, and $K_{\mathrm{s}}$ data, as well as optical H$\alpha$, [\ion{S}{ii}]$\lambda6717,31$ data from the Danish telescope at La Silla to characterize the flow. From their analysis these authors find what seems to be a southern counterpart to HH 216 along a pillar south-east of HH 216, which in terms of radial velocity is blueshifted with respect to the redshifted HH object. They also find the counterpart to consist of several knots. However, their data does not allow the identification of a possible source of the HH flow. Here, HH~216 is seen as a bright bow shock-shaped feature in all of the emission line maps shown in Section 3. In the [\ion{O}{ii}] emission lines, HH~216 shows two distinct components (Figure~\ref{fig:spectra}), where the double-peaked line profile is a consequence of the bow-shock geometry of HH~216 \citep{caratti09}. 

In this work we are able to expand on the near-IR analysis of \cite{andersen04} by including WIRCam narrow-band Br$\gamma$ and H$_{2}$ data and therefore present a more detailed study of the HH 216 counterpart in terms of molecular and ionized emission lines within the $K$-band filter. We combine our narrow-band near-IR images with the SITELLE optical nebular emission line maps to analyze the structure and kinematics of the counterflow of HH~216.

\begin{figure*}[!t]
    \centering
    \includegraphics[width=0.45\linewidth]{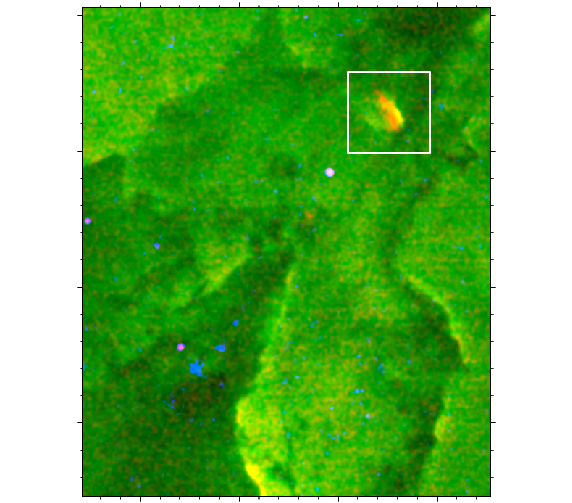}
    \includegraphics[width=0.45\linewidth]{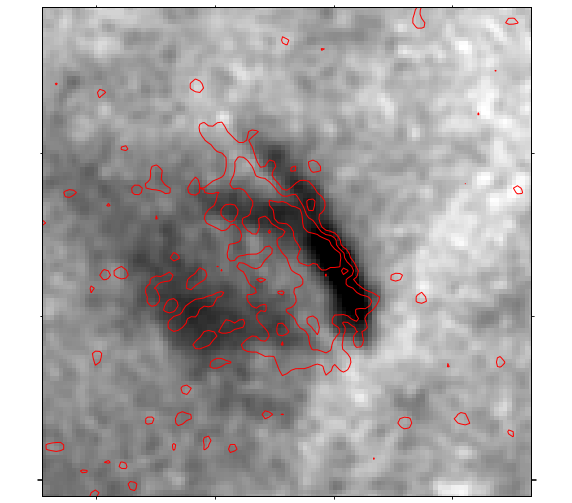}
    \caption{Left panel is a three color image of the flux for the three [\ion{O}{ii}] radial velocity components: red is the [\ion{O}{ii}]$\lambda$3729 redshifted component, green is the [\ion{O}{ii}]$\lambda$3729 main component, and blue is the [\ion{O}{ii}]$\lambda$3726 blueshifted component. Right panel is a zoom on the white box from the left panel showing the main radial velocity component in grayscale with the redshifted component as contours.}
    \label{fig:oii_velocity}
\end{figure*}

Figure~\ref{hh216} shows the H$\alpha$ and [\ion{S}{ii}]$\lambda$6717 flux maps of the region containing the HH object, as well as the H$\alpha$ radial velocity map, scaled from -45 to 40 km s$^{-1}$. The outline of the pillar is best recognized in the H$\alpha$ map, while the emission line knots described in \cite{andersen04} can be clearly identified in the [\ion{S}{ii}] map. The sequence of knots is parallel to the pillar body, which -- just as the main central pillars -- point back towards the ionizing stars of NGC 6611. By eye we identify seven emission line knots, the spatial distribution of which is discussed at the end of this section.

The radial velocity map in the right panel of Figure~\ref{hh216} shows the emission line knots as being blueshifted with respect to the HH object in the upper right corner of the map. Furthermore, redshifted material can be identified just above the pillar tip in line with the blueshifted knots. As already mentioned in \cite{andersen04}, the observations suggest that HH 216 is the bow shock of a bipolar jet. The SITELLE radial velocity map delivers solid evidence in support of this, showing a clear, coherent bipolar geometry with a total extent of 2.8 arcminutes ($\sim$ 1.8 pc at a distance of 2 kpc). 

\begin{figure*}[!t]
\centering
\includegraphics[scale=0.6, trim= 3cm 0cm 4cm 0cm]{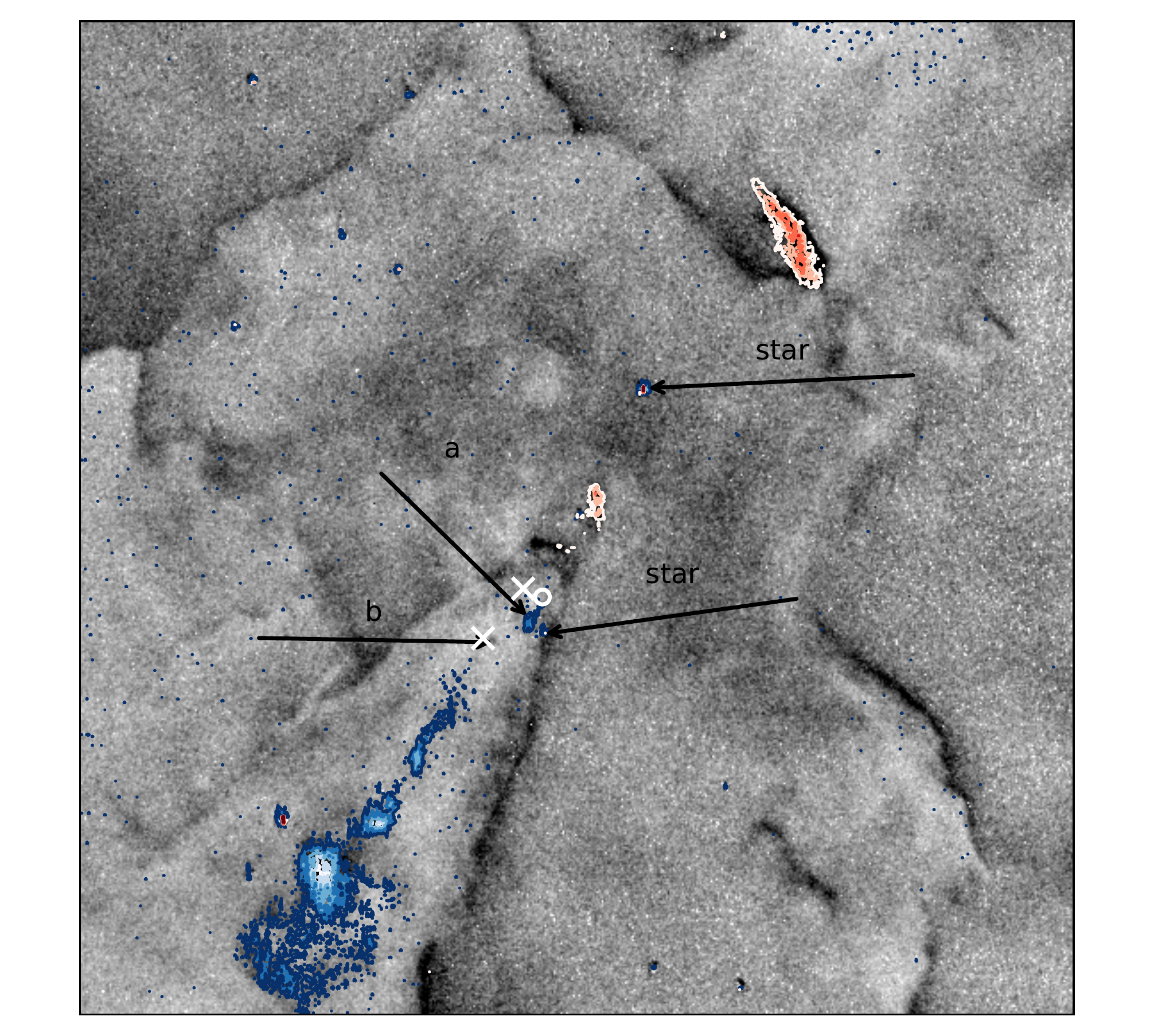}
\caption{[\ion{S}{ii}]$\lambda$6717 integrated line map with contours from the H$\alpha$ radial velocity map. Blue contours range from -50 to -15 km s$^{-1}$ and red contours from 15 to 50 km s$^{-1}$. The white crosses correspond to the positions of HH-N and HH-S (upper and lower cross, respectively), the white circle indicates the location of an infrared point source (see Figure~\ref{nir}), while the black arrows indicate the locations of knots $a$ and $b$, as well as of two stellar sources.}
\label{SII_vel}
\end{figure*}

\begin{figure*}[!t]
\centering
\includegraphics[width=0.45\linewidth]{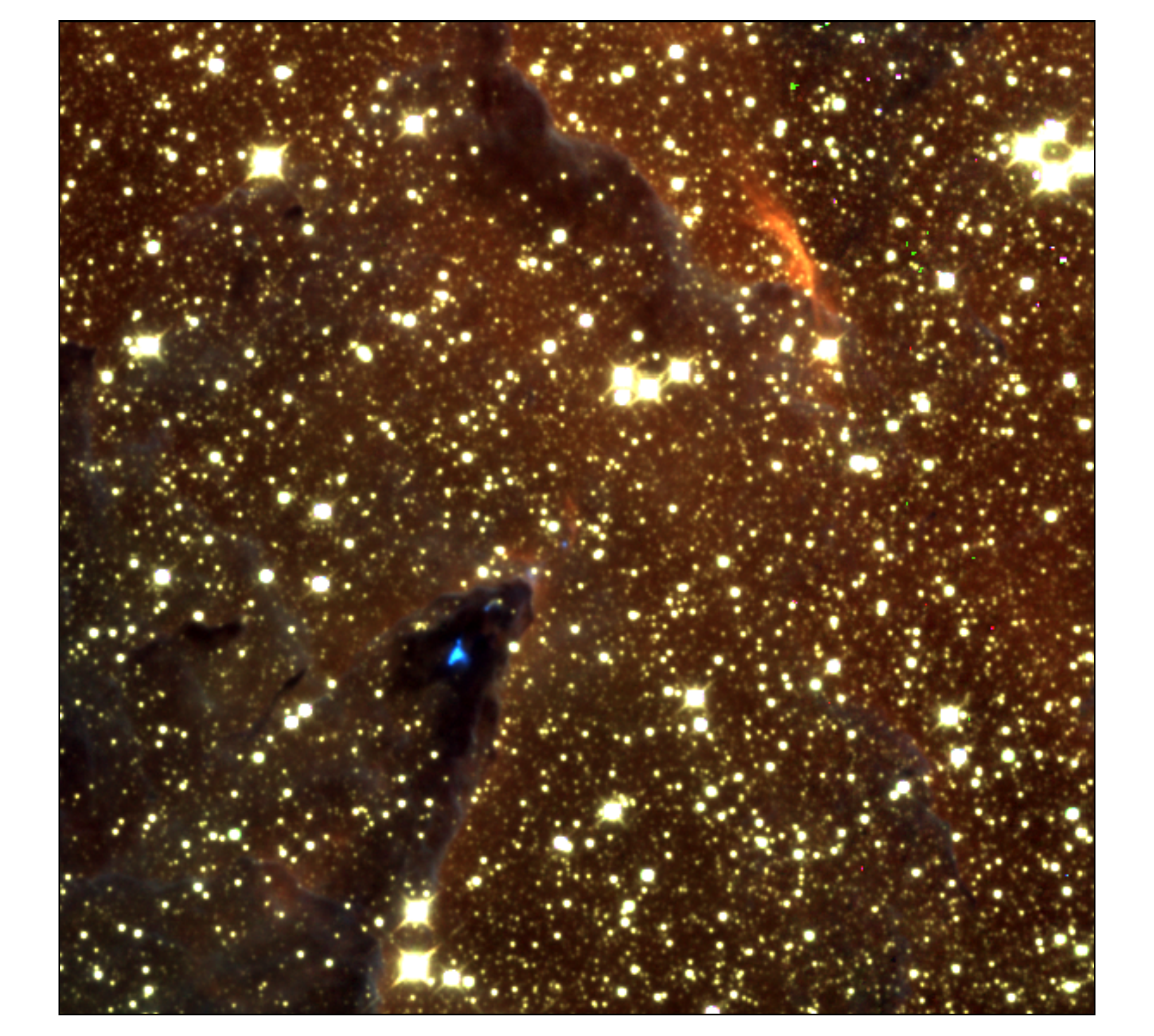}
\includegraphics[width=0.45\linewidth]{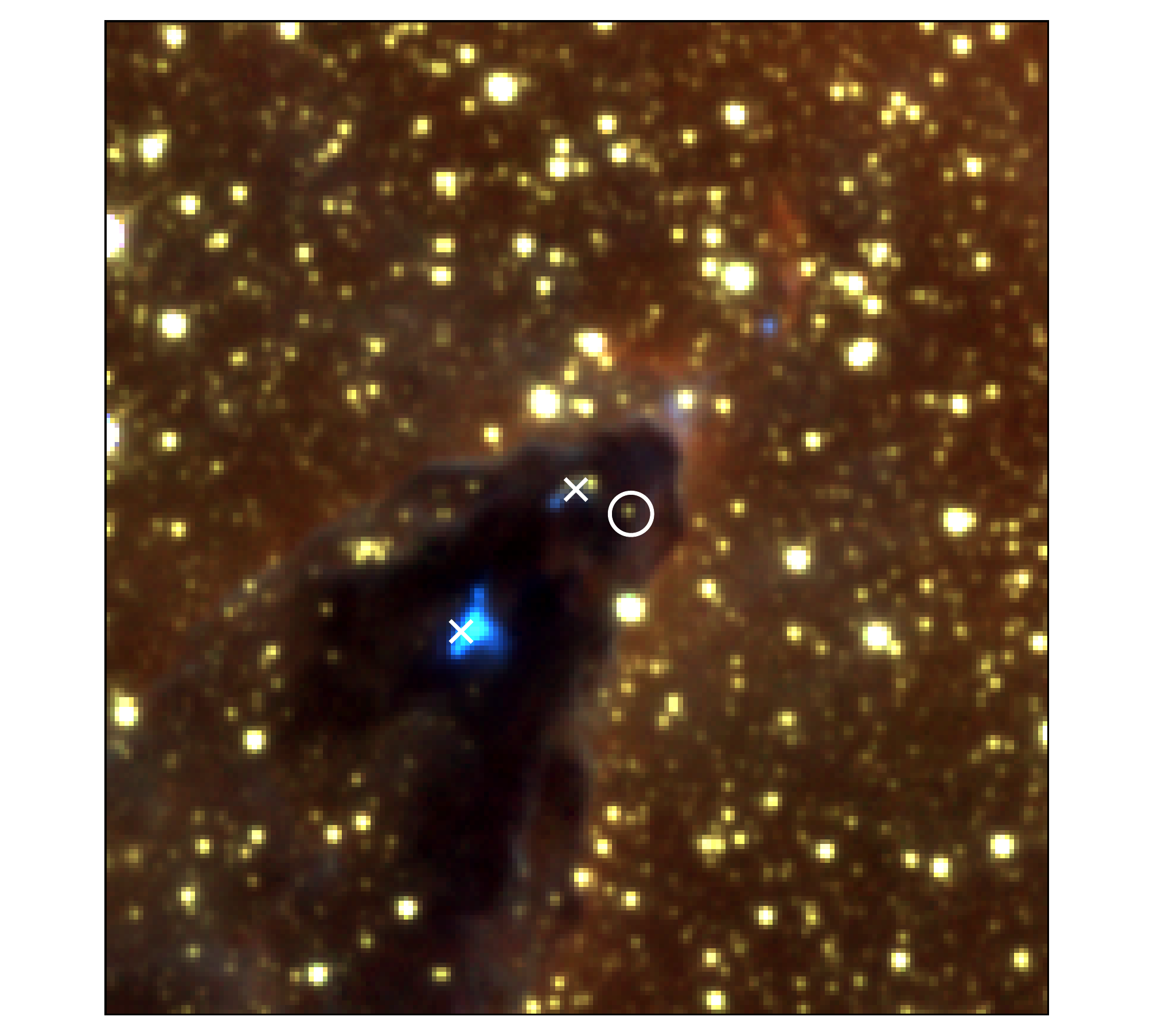}
\caption{RGB composites of HH216 (red is Br$\gamma$, green is K$_{s}$, and blue is H$_{2}$). The left panel is a zoom of the right panel onto the pillar tip. The white crosses correspond to the positions of HH-N and HH-S (upper and lower cross, respectively), while the white circle is centered on a point source at the tip of the pillar identified in the Br$\gamma$ image.}
\label{nir}
\end{figure*}

\citet{Meaburn1982} find that HH~216 (M16-HH1 in their work) has two dominant radial velocity components separated by about 80~km/s, with which our analysis agrees very well (70~km/s). \citet{meaburn90} find that HH~216 can be traced at radial velocities up to 150~km/s in the [\ion{O}{iii}]$\lambda$5007 line at a spectral resolution of more than 30'000. Their main component is typically at 10-20~km/s while the high-speed component of HH~216 is mostly around 80-90~km/s (see their Figures 5 and 6). Because of the lower resolution of the SN2 and SN3 cubes, we do not detect these high radial velocity components of HH~216 in the H$\alpha$ or [\ion{S}{ii}] maps. As described in Section \ref{sec:vel}, we do, however, detect multiple radial velocity components of HH~216 in the higher-resolution SN1 cube covering the [\ion{O}{ii}] lines, which trace the flow to a peak relative radial velocity of $\sim$ 70~km/s, in very good agreement with the work of \citet{meaburn90}. Figure \ref{fig:oii_velocity} shows the three radial velocity components of the [\ion{O}{ii}] doublet identified in the SN1 spectral cube (with these lines being much fainter than the H$\alpha$ line, the lower S/N is clearly noticeable when compared to the H$\alpha$ kinematics shown in the right panel of Figure~\ref{hh216}). The bow shock of HH~216 clearly shows spatially distinct features: the redshifted component is slightly to the South-East of the main component (see right panel of Figure \ref{fig:oii_velocity}), which seems to agree with Figure 7 of \citet{meaburn90}.

\cite{indeb07} identify two highly-extincted Class I YSOs at the tip of the pillar-like structure the bipolar jet is emerging from, and suggest that these are probably not high-luminosity sources. One of these two sources, HH-N, is at the very tip of the pillar, while HH-S is a little further south (the nomenclature HH-N and HH-S is as in \citealt{indeb07}). The source located more towards the tip of the pillar, HH-N, is associated with maser activity \citep{healy04}. To better visualize the geometry of the HH~216 counterflow, in Figure~\ref{SII_vel} we plot the contours of the radial velocity map onto the [\ion{S}{ii}] map, together with the coordinates of HH-N and HH-S (white crosses). Figure~\ref{SII_vel} suggests that the red lobe emerges at the tip of the pillar, while the blue lobe is detected south of HH-N at the pillar tip, but north of the source HH-S further down in the pillar, indicating that HH-S is an unlikely source for the flow. However, the near-infrared images reveal a slightly different picture (Figure~\ref{nir}). HH-S is associated with a bright H$_{2}$ knot, which corresponds to knot \textit{b} in Figure~\ref{hh216}, while two faint point sources are located at the top of the pillar. One of these is very close to the coordinates given in \cite{indeb07} for HH-N and is the likely NIR counterpart of HH-N, and a fainter H$_{2}$ knot is detected just $\sim$1.5" South of it. The second NIR point source is located towards the right edge of the pillar (black and white circles in Figure~\ref{hh216} and \ref{nir}, respectively), just above the beginning of the blueshifted lobe, corresponding to the location of knot \textit{a}. Our data set does not allow a conclusive answer as to which of the possible objects is the definite source of the HH 216 flow, however, based on the above findings we suggest that neither HH-N nor HH-S are powering the HH216 flow, and that the detected point source to the right of HH-N is the more likely source of the flow. 

Several authors describe knots in HH objects and suggest that these jet knots are likely related to episodic accretion bursts or fluctuations of the accretion rate of the source star, which lead to separations between the knots of the order of the accretion burst frequency \citep{reipurth85} or accretion rate fluctuations \citep{ioannidis12}. The knots of the blue lobe have separations between $\sim$ 0.07 - 0.19 pc (at a distance of 2 kpc), and, together with observed relative radial velocities of the order of $\sim$ 70 km s$^{-1}$, we find an ejection timescale of the order of 1-3 kyr. This timescale agrees with values found for outflows in Serpens and Aquila (e.g., \citealt{ioannidis12}), and is slightly lower than the lower limit of predicted FU Ori outburst intervals of 5-50 kyr \citep{scholz13}. With an extent of $\sim$ 0.82 pc of the blue lobe, the derived dynamical timescale of the jet is about $1\times10^{4}$yr, which corresponds to the typical timescales of YSO jets. Finally, we note that while the SITELLE data is insufficient for further investigations, the emission line knots of the blueshifted lobe show a weak S-shaped morphology, which could be an indication for jet precession \citep{reipurth01}. Hence, in its morphology, appearance, and driving source, HH~216 is similar to the iconic HH~46-47 system \citep[see e.g.,][]{doptia82,reipurth91}.

\section{Mass-loss rate of the Spire}
\label{sec:spire}

\begin{figure}
\includegraphics[scale=0.6]{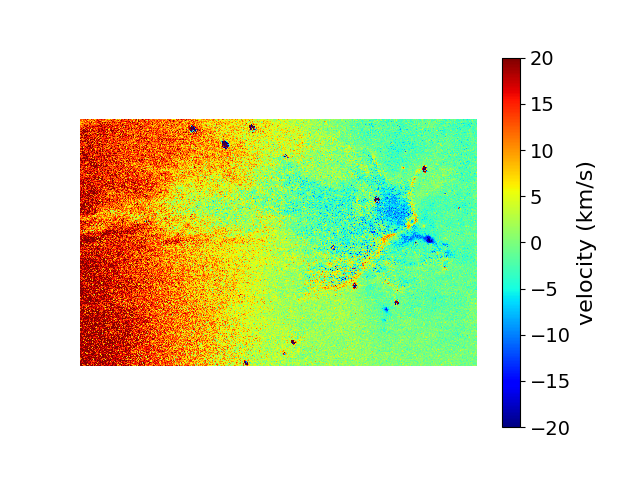}
\caption{H$\alpha$ radial velocity map of the North-Eastern pillar (the Spire).}
\label{spire_vel}
\end{figure}

The SITELLE data also cover a second major pillar, which is located approximately 5.5\arcmin~(3.2 pc) North-East of the well-known three pillars at the center of the image (see Figure~\ref{rgb}). This pillar, dubbed the "Spire", has a narrow, ionized tip which quickly widens into a fringed pillar body. Just above its tip, a small ionized globule with a tadpole-like tail can be identified, which is suggested to be a HH object \citep{meaburn86}. \cite{healy04} identify several water masers at the tip of the Spire. As for the other pillars in the region (see Figure~\ref{spire_vel}), the radial velocity map of the Spire shows the pillar material being blueshifted with respect to the ambient matter, and the globule at the tip showing the bluest relative radial velocities of the region ($v\sim$-20 km s$^{-1}$).

\begin{figure*}
\centering
\includegraphics[scale=0.5,trim=4cm 0cm 0cm 0cm]{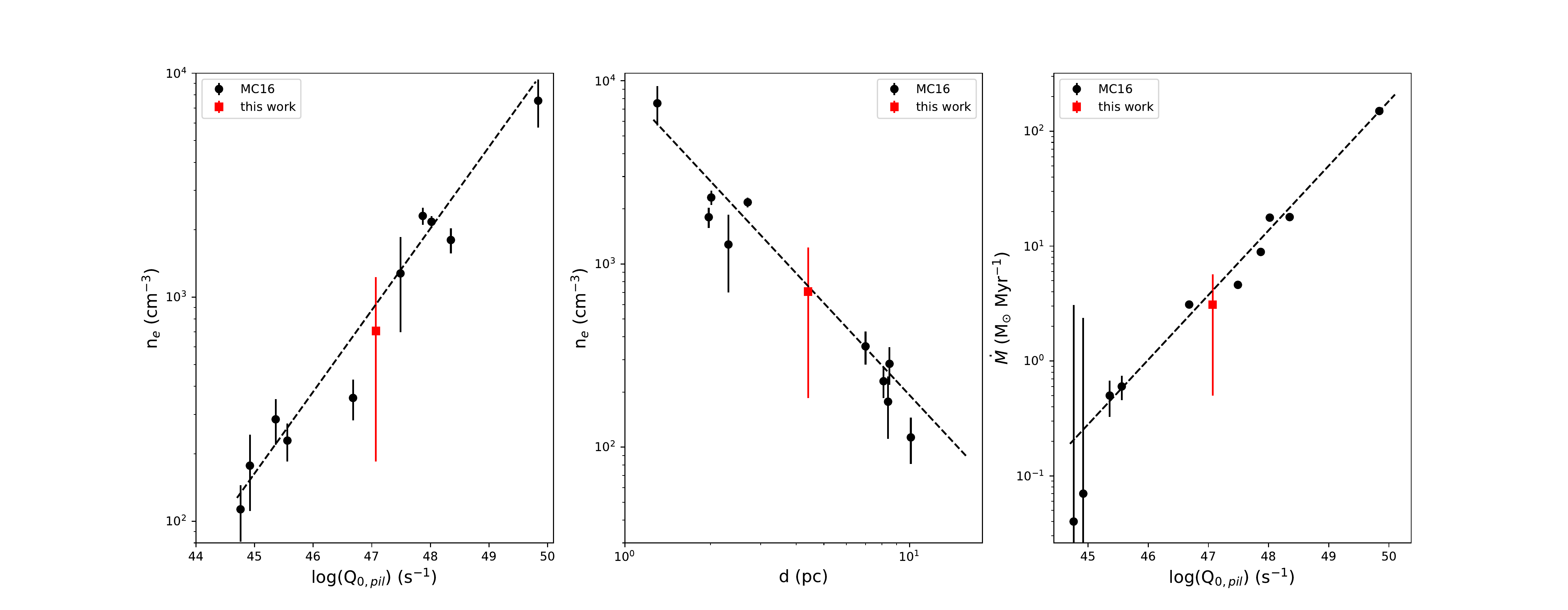}
\caption{Electron density $n_{e}$ as a function of the incident ionizing photon flux $Q_{0,pil}$ (left) and as a function of distance from the main ionizing sources (middle). The right panel shows the mass-loss rate as a function of $Q_{0,pil}$. The black data points and best fit relations are from MC16 \citep{pillars}, while the red square corresponds to the values obtained with the SITELLE data presented in this work.}
\label{mdot}
\end{figure*}

With MUSE optical integral field data, \cite{pillars} (henceforth referred to as MC16) computed the mass-loss rate due to photoevaporation of pillar-like structures in the Carina Complex, M~16, and NGC~3603. A tight correlation between the mass-loss rate and the incident ionizing photon flux $Q_{0}$ was found, showing that the mass-loss rate increases with increasing $Q_{0}$ and delivering a quantitative analysis of ionization feedback from massive stars. MC16 also found correlations for the electron density versus $Q_{0}$, and the electron density versus the projected distance to the ionizing massive stars, showing that only the densest pillars are able to survive in the regions immediately close to the ionizing massive stars.

Here, we exploit the larger spatial coverage of M~16 provided by the SITELLE data to compute the mass-loss rate of the Spire and expand the sample of pillars used in MC16. For this, we use the same methodology as in MC16. We first compute the electron density n$_{e}$ within a $\sim$ 3" circular aperture at the ionized tip of the Spire to be $\approx$ 706 cm$^{-3}$ using the density sensitive ratio of the [\ion{S}{ii}] lines as in MC15. From the radial velocity map, we extract a velocity of the photoevaporative flow of $\sim$ 8 km s$^{-1}$, value comparable to the values found for photoevaporative flows of similar structures (MC15, \citealt{pillars}). We compute the (projected) distance of the Spire to the main ionizing source in NGC 6611 (HD 168076) and obtain d $\sim$ 4.4 pc. With a radius of the very tip of the pillar of $\approx$ 0.07 pc and an ionizing photon flux $\mathrm{log(Q_{\mathrm{0}})}\sim$49.87 for NGC 6611 (MC16), the ionizing photon flux perceived at the pillar tip at a distance of 4.4 pc from the ionizing source is $\mathrm{log(Q_{\mathrm{0,pil}})}\sim$ 47.07. Inserting the values for velocity, pillar radius, and density into Equation~1 in MC16, we then obtain a mass-loss rate of the Spire of $\dot{M}\simeq\pi r^{2}m_{H}n_{H}v$ $\sim$ 3 M$_{\odot}$ Myr$^{-1}$.

Because of the significant noise in the map, we allow a generous 50\% error for n$_{\mathrm{e}}$, while for the projected distance we assume a generous 20\% measurement error. The obtained values of the mass-loss rate $\dot{M}$, photon flux at the pillar tip $Q_{\mathrm{0,pil}}$, and density of the Spire, as compared to those obtained in MC16, are shown in Figure~\ref{mdot}. The mass-loss rate derived for the Spire from the SITELLE data therefore agrees very well with the relation previously found in MC16. The Spire is yet another example of a dense molecular cloud structure being shaped and photoevaporated by the feedback of the nearby massive stars. As mentioned in MC16, with molecular CO data one could potentially compute the mass of the Spire, and, together with the mass-loss rate, estimate the lifetime of the pillar.

\section{Summary and conclusions}
\label{sec:ccl}

Here we present SITELLE Fourier transform spectroscopy of the iconic star-forming region M~16, also known as the Eagle Nebula. With the covered wavelength ranges, we perform a systematic analysis of the performance and data reduction of SITELLE by comparing the SITELLE integrated line maps to those obtained with the optical integral field spectrograph MUSE mounted on the VLT and other spectroscopic and narrow-band imaging data. We find very good agreement overall between the various data sets and confirm the power of an instrument like SITELLE for observing wide nebulae.

We exploit the emission line coverage to derive the oxygen abundance of M~16 while taking the dependence of the abundance tracing line ratios on the ionization structure of the nebula into consideration. 

We confirm the presence of a counterflow to the Herbig-Haro object HH~216. The flow presents a clear bipolar structure that spans about 1.8 pc, and we tentatively identify the likely driving source of the flow as a star within the tip of a pillar-like structure located south-east of HH~216. While HH~216 is a bow-shock associated with the entire flow, the blueshifted counterpart presents itself as a series of emission line knots along the surface of above-mentioned pillar, the spacing of which indicate episodic accretion bursts on a timescale of about 1-3~kyr over a total lifetime of about 10~kyr.

Finally, we measure the mass-loss rate $\dot{M}$ due to photo-evaporation of the eastern pillar (the Spire) to test the correlation between $\dot{M}$ and the ionizing feedback from nearby massive stars presented in \cite{pillars}. We find that the mass-loss rate of the Spire fits very well with said correlation, therefore augmenting the sample of the \cite{pillars} pillars (all observed with MUSE) with data from a different instrument. 

We conclude that SITELLE is a high-performing instrument in terms of observations of H{\sc ii} regions. It offers an unprecedented large FOV of 11\arcmin$\times$11\arcmin\ in combination with a wavelength coverage which includes all the main collisionally excited and recombination lines typically used in nebular studies (i.e., [\ion{O}{ii}]$\lambda$3726,29, H$\beta$, [\ion{O}{iii}]$\lambda$4959,5007, [\ion{N}{ii}]$\lambda$6548,84, H$\alpha$, and [\ion{S}{ii}]$\lambda$6717,31). For these lines, SITELLE can deliver spatial and kinematical information, with a resolving power from a few 100s to a few 1000s. This instrument proves therefore to be ideal for large-scale H{\sc ii} region surveys aimed at nebular analyses which require simultaneous spatial and spectral coverage.

\begin{acknowledgements}
The authors want to thank Laurent Drissen, P.I. of SITELLE, and Thomas Martin for their support and for discussion that helped improve the paper. The authors also want to thank the referee, John Meaburn, for his comments that help improve the paper.

Part of this work was supported by the Royal Society of New Zealand through AM's Marsden grant.

Support for this work was provided by NASA through an award issued by the Jet Propulsion Laboratory and the California Institute of Technology.

Based on observations obtained with SITELLE, a joint project between Université Laval, ABB-Bomem, Université de Montréal, and the CFHT with funding support from the Canada Foundation for Innovation (CFI), the National Sciences and Engineering Research Council of Canada (NSERC), Fond de Recheche du Québec - Nature et Technologies (FRQNT) and CFHT.
\end{acknowledgements}

\bibliographystyle{aa}
\bibliography{sitelle_refs}

\begin{appendix}

\section{Fit parameters}
Table \ref{tab:fit} lists all the parameters of the fits. The initial guesses and the authorized ranges we used for each parameter, when applicable, are also indicated.

\begin{table*}[!t]
  \centering
  \caption{Initial guess and limits used for each parameter in the fit of the SN1, SN2 and SN3 spectral cubes.}
  \begin{tabular}{c c | c c | c c | c c}
    \hline
    \multicolumn{2}{c|}{Component}  & \multicolumn{2}{c|}{Position} & \multicolumn{2}{c|}{Amplitude} & \multicolumn{2}{c}{Width} \\
     & & \multicolumn{2}{c|}{(cm$^{-1}$)} & \multicolumn{2}{c|}{(erg/cm$^{2}$/s/{\AA})} & \multicolumn{2}{c}{(cm$^{-1}$)} \\
     & & Initial & Limits & Initial & Limits & Initial & Limits \\
    \hline
    \hline
    \multirow{7}{*}{SN1} & Continuum &  \multicolumn{2}{c|}{N/A} &   $1\times 10^{-16}$ & $>0$ & \multicolumn{2}{c}{N/A} \\
	& [\ion{O}{ii}]~3726 (red)       &   26800 & 26798.1-26801.9 &   $1\times 10^{-16}$ & $>0$ &   2 & $>$0 \\
	& [\ion{O}{ii}]~3726 (main)      &   26808 & 26806.1-26809.9 &   $5\times 10^{-16}$ & $>0$ &   2 & $>0$ \\
    & [\ion{O}{ii}]~3726 (blue)      &   26816 & 26814.1-26817.9 &   $1\times 10^{-16}$ & $>0$ &   2 & $>0$ \\
    & [\ion{O}{ii}]~3729 (red)       &   26820 & 26818.1-26821.9 &   $1\times 10^{-16}$ & $>0$ &   2 & $>0$ \\
    & [\ion{O}{ii}]~3729 (main)      &   26828 & 26826.1-26829.9 &   $5\times 10^{-16}$ & $>0$ &   2 & $>0$ \\
    & [\ion{O}{ii}]~3729 (blue)      &   26836 & 26834.1-26837.9 &   $1\times 10^{-16}$ & $>0$ &   2 & $>0$ \\
	\hline
    \multirow{4}{*}{SN2} & Continuum &  \multicolumn{2}{c|}{N/A} &   $1\times 10^{-17}$ & $>0$ & \multicolumn{2}{c}{N/A} \\
    & [\ion{O}{III}]~5007            &   19970 &   19966.5-19974 &   $1\times 10^{-16}$ & $>0$ &  10 & $>0$ \\
    & [\ion{O}{III}]~4960            &   20164 &     20157-20171 &   $4\times 10^{-17}$ & $>0$ &  10 & $>0$ \\
    & H$\beta$                       & 20570.5 &     20566-20575 &   $1\times 10^{-16}$ & $>0$ &  10 & $>0$ \\
    \hline
	\multirow{6}{*}{SN3} & Continuum &  \multicolumn{2}{c|}{N/A} &   $1\times 10^{-17}$ & $>0$ & \multicolumn{2}{c}{N/A} \\
    & [\ion{S}{II}]~6731             &   14852 &     14848-14859 &   $3\times 10^{-17}$ & $>0$ & 3.7 & $>0$ \\
    & [\ion{S}{II}]~6717             &   14885 &     14879-14891 & $3.5\times 10^{-17}$ & $>0$ &   5 & $>0$ \\
    & [\ion{N}{ii}]~6584             &   15185 &     15180-15190 & $7.5\times 10^{-17}$ & $>0$ &   5 & $>0$ \\
    & H$\alpha$                      &   15233 &     15230-15236 &   $4\times 10^{-16}$ & $>0$ &   5 & $>0$ \\
    & [\ion{N}{ii}]~6548             &   15267 &     15262-15273 &   $5\times 10^{-17}$ & $>0$ &   5 & $>0$ \\
    \hline
  \end{tabular}
  \label{tab:fit}
\end{table*}

\section{Flux maps}

\begin{figure*}[h]
\centering
\subfloat[H$\alpha$]{\includegraphics[scale=0.38]{flux_Halpha.png}}
\hspace{0.05in}
\subfloat[H$\beta$]{\includegraphics[scale=0.38]{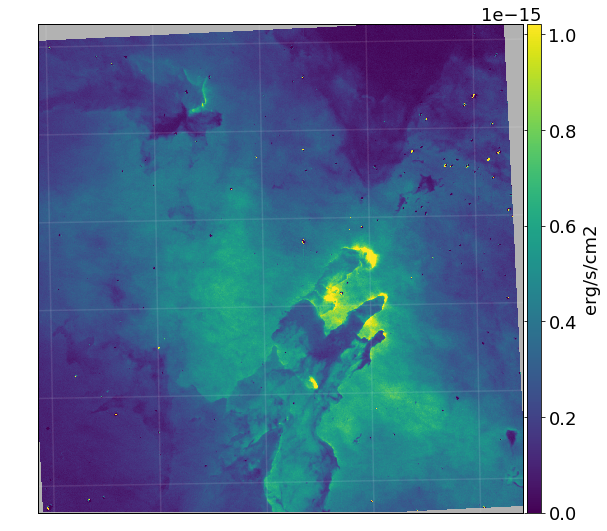}}\\
\subfloat[ $\rm{[\ion{O}{iii}]\lambda 4960}$ ]{\includegraphics[scale=0.38]{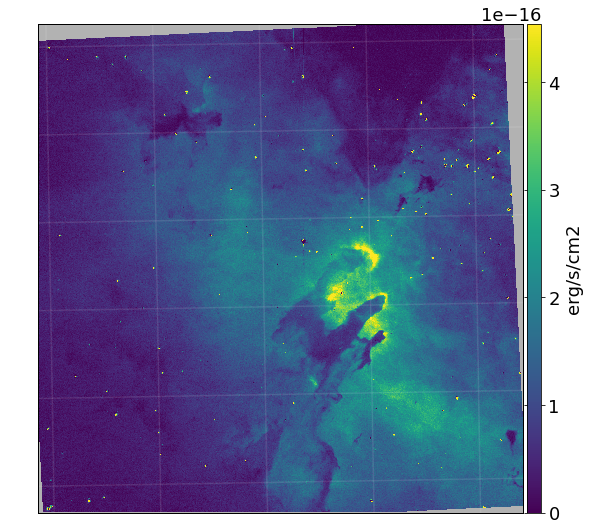}}
\hspace{0.05in}
\subfloat[ $\rm{[\ion{O}{iii}]\lambda 5007}$ ]{\includegraphics[scale=0.38]{flux_OIII5007.png}}
\caption{Emission line flux maps (in $\rm{erg/s/cm^2}$) for the hydrogen and the [\ion{O}{iii}] lines.}
\label{fig:fluxmaps1}
\end{figure*}

\begin{figure*}[h]
\centering
\subfloat[ $\rm{[\ion{N}{ii}]\lambda 6548}$ ]{\includegraphics[scale=0.38]{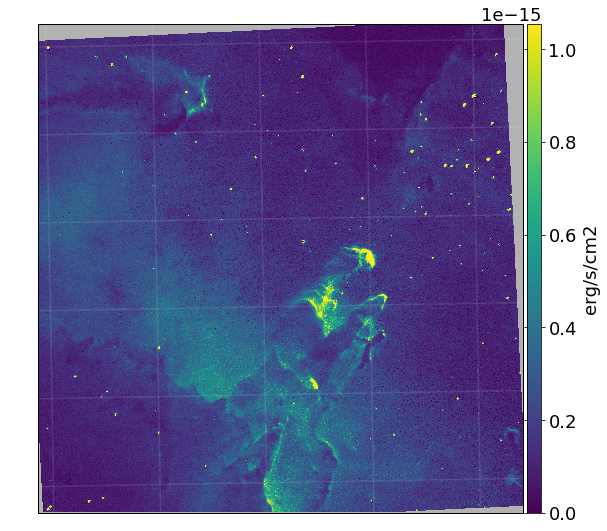}}
\hspace{0.05in}
\subfloat[ $\rm{[\ion{N}{ii}]\lambda 6584}$ ]{\includegraphics[scale=0.38]{flux_NII6584.png}} \\
\subfloat[ $\rm{[\ion{S}{ii}]\lambda 6717}$ ]{\includegraphics[scale=0.38]{flux_SII6717.png}}
\hspace{0.05in}
\subfloat[ $\rm{[\ion{S}{ii}]\lambda 6731}$ ]{\includegraphics[scale=0.38]{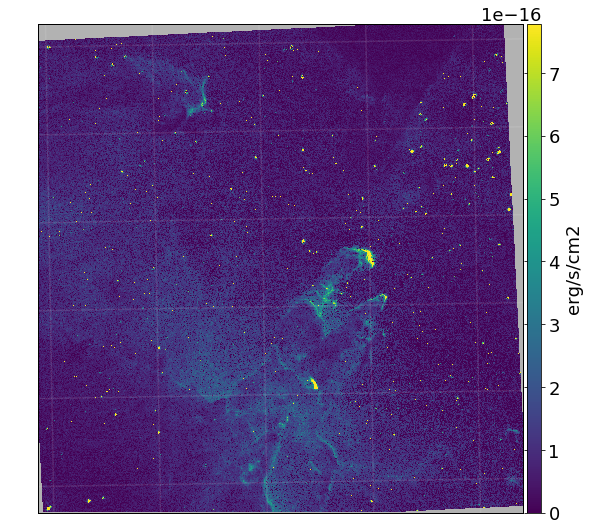}} \\
\caption{Emission line flux maps (in $\rm{erg/s/cm^2}$) for the [\ion{N}{ii}] and [\ion{S}{ii}] lines.}
\label{fig:fluxmaps2}
\end{figure*}

\begin{figure*}[h]
\centering
\subfloat[ $\rm{[\ion{O}{ii}]\lambda 3726}$ ]{\includegraphics[scale=0.34]{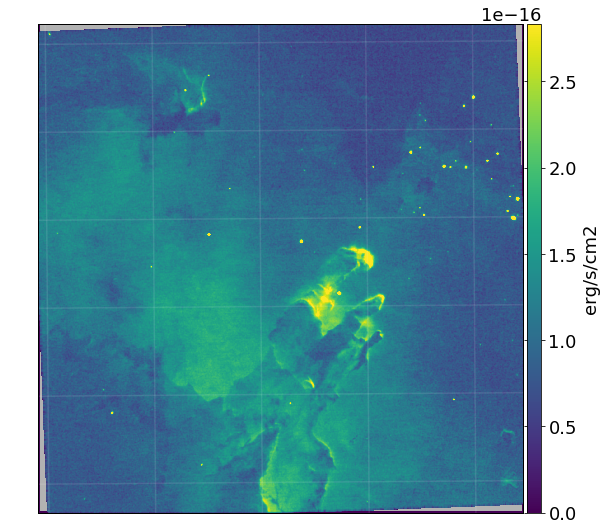}}
\hspace{0.05in}
\subfloat[ $\rm{[\ion{O}{ii}]\lambda 3729}$ ]{\includegraphics[scale=0.34]{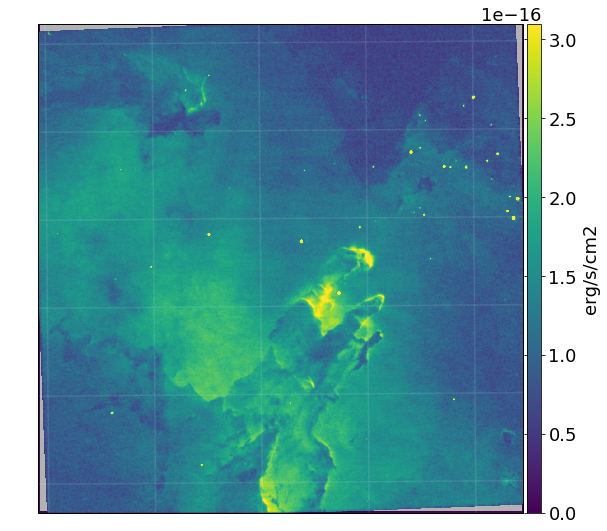}} \\
\subfloat[ $\rm{[\ion{O}{ii}]\lambda 3726, red}$ ]{\includegraphics[scale=0.34]{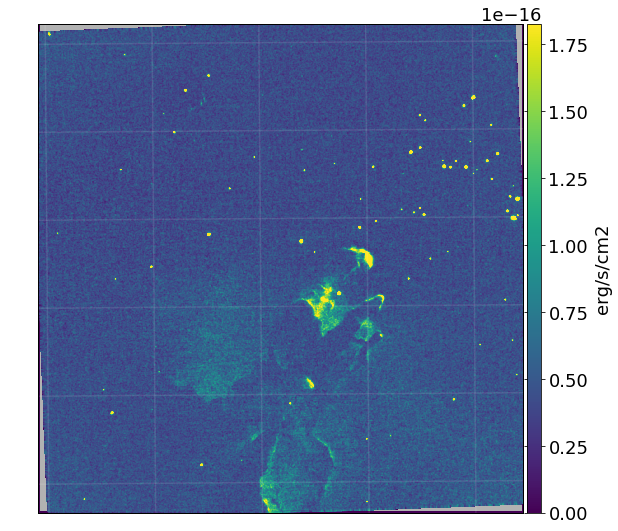}}
\hspace{0.05in}
\subfloat[ $\rm{[\ion{O}{ii}]\lambda 3729, red}$ ]{\includegraphics[scale=0.34]{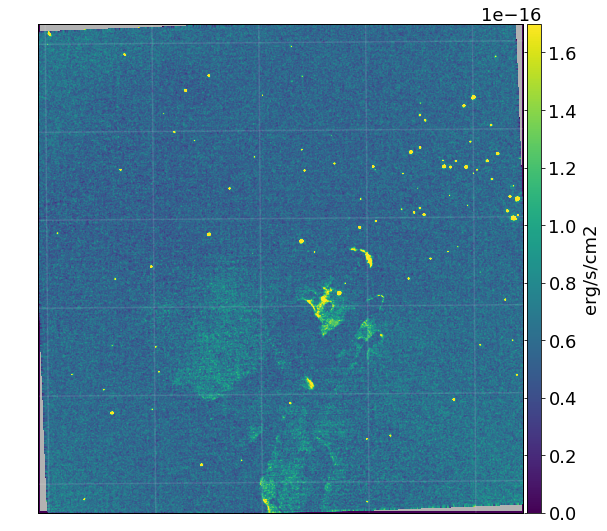}} \\
\subfloat[ $\rm{[\ion{O}{ii}]\lambda 3726, blue}$ ]{\includegraphics[scale=0.34]{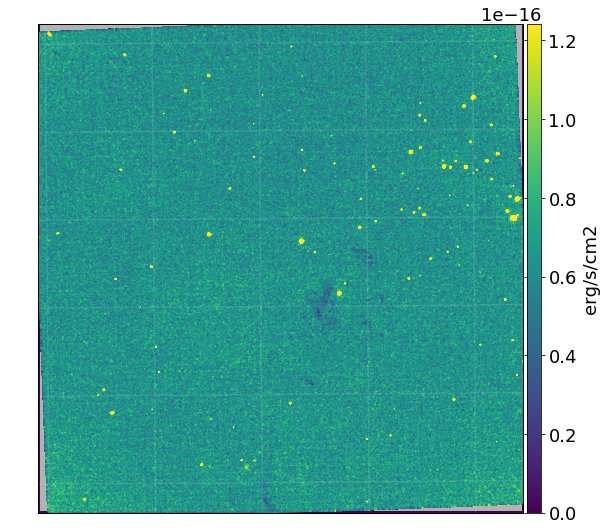}}
\hspace{0.05in}
\subfloat[ $\rm{[\ion{O}{ii}]\lambda 3729, blue}$ ]{\includegraphics[scale=0.34]{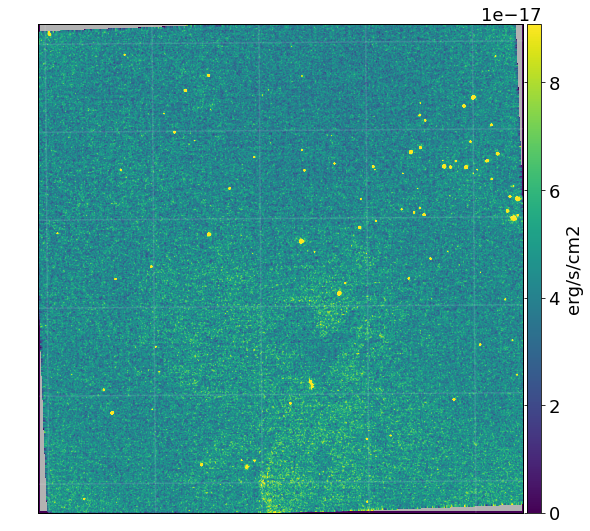}} \\
\caption{Emission line flux maps (in $\rm{erg/s/cm^2}$) for the [\ion{O}{ii}]$\lambda$3726 and [\ion{O}{ii}]$\lambda$3729 lines. The images have been smoothed with a Gaussian convolution ($\sigma=2$ pixels).}
\label{fig:fluxmaps3}
\end{figure*}

Figure \ref{fig:fluxmaps1} shows the flux maps for the hydrogen and [\ion{O}{iii}] emission lines, Figure \ref{fig:fluxmaps2} for the [\ion{N}{ii}] and [\ion{S}{ii}] emission lines, and Figure \ref{fig:fluxmaps3} for the [\ion{O}{ii}] emission lines.

\end{appendix}

\end{document}